

\mag 1200

\input amstex
\input amsppt.sty

\overfullrule 0pt

\hsize 5.25in
\vsize 7in

\leftheadtext{V\. Tarasov and A\. Varchenko}
\rightheadtext{Jackson Integral Representations $\ldots$ QKZ Equation}

\let\bls\baselineskip
\def\vsk#1>{\vskip#1\bls} \def\vv#1>{\vadjust{\vsk#1>}}
\def\vvn#1>{\vadjust{\nobreak\vsk#1>\nobreak}}

\let\vp\vphantom  \let\<\negthickspace
\let\nl=\newline \let\nt\noindent \let\cl\centerline
\def\nn#1>{\noalign{\vskip #1pt}} \def\NN#1>{\openup#1pt}

\let\alb\allowbreak \def\({\allowbreak(} \def\alh{\hfil\alb\hfilneg}
\def\ald{\noalign{\alb}} \let\alds\allowdisplaybreaks

\def\Cup{\bigcup\limits} \def\Cap{\bigcap\limits}
\let\Sum\sum \def\sum{\Sum\limits} \def\Plus{\bigoplus\limits}
\let\Prod\prod \def\prod{\Prod\limits} \let\Int\int \def\int{\Int\limits}
\def\tsum{\mathop{\tsize\Sum}\limits}

\let\o\circ \let\x\times \let\ox\otimes
\let\sub\subset 
\let\le\leqslant \let\ge\geqslant
\let\der\partial \let\D\nabla \let\8\infty
\let\bra\langle \let\ket\rangle
\let\und\underbrace \let\plus\dotplus

\let\al\alpha
\let\bt\beta
\let\gm\gamma \let\Gm\Gamma
\let\dl\delta \let\Dl\Delta
\let\la\lambda \let\La\Lambda
\let\si\sigma
 \let\phi\varphi
\let\om\omega \let\Om\Omega

\def\C{\Bbb C}

\def\T{\Bbb T}
\def\Z{\Bbb Z}

\def\lsym#1{#1\alb\ldots#1\alb}
\def\lc{\lsym,}  \def\lx{\lsym\x} \def\lox{\lsym\ox}
\newbox\dibox

\def\hleft#1:#2{\setbox\dibox\hbox{$\dsize #1\quad$}\rlap{$\dsize #2$}
 \kern-2\wd\dibox\kern\displaywidth}
\def\hright#1:#2{\setbox\dibox\hbox{$\dsize #1\quad$}\kern-\wd\dibox
 \kern\displaywidth\kern-\wd\dibox\llap{$\dsize #2$}}

\newbox\sectbox
\def\sect#1 #2\par{\removelastskip\vskip.8\bls
 \vtop{\bf\setbox\sectbox\hbox{#1} \parindent\wd\sectbox
 \ifdim\parindent>0pt\advance\parindent.5em\fi\item{#1}#2\strut}%
 \nointerlineskip\nobreak\vtop{\strut}\nobreak\vskip-.6\bls\nobreak}

\def\E(#1){\mathop{\hbox{\rm End}\,}(#1)} \def\im{\mathop{\hbox{\rm im}\;\!}}
\def\Li{\mathop{\hbox{\rm Li}_2}} \def\id{\hbox{\rm id}}
\def\vst#1{{\lower2.1pt\hbox{$\bigr|_{#1}$}}}
\def\1{^{-1}} \def\0{^{\vp1}}
\def\tagg"#1"{\tag"\rlap{(#1)}"}

\def\Text#1{\noalign{\alb\vsk>\normalbaselines\vsk->\vbox{\nt #1\strut}%
 \nobreak\nointerlineskip\vbox{\strut}\nobreak\vsk->\nobreak}}
\def\Remark#1{\remark{{\rm #1} Remark}} \def\Example{\example{Example}}
\def\Proof#1.{\demo{Proof #1}} \let\endproof\enddemo

\def\egv/{eigenvector} \def\eva/{eigenvalue}
\def\eq/{equation} \def\eqt/{equivalent} \def\stt/{statement}
\def\lhs/{the left hand side} \def\rhs/{the right hand side}
\def\Rm/{$R$-matrix} \def\Rms/{$R$-matrices}
\def\rep/{representation} \def\YB/{Yang-Baxter \eq/}

\def\fps/{formal power series} \def\cc/{compatibility condition}
\def\dfl/{differential} \def\dip/{\dfl/ polynomial} \def\perm/{permutation}
\def\dif/{difference} \def\deq/{\dif/ \eq/} \def\dc/{discrete}
\def\wt/{weight} \def\m/{moqule} \def\hw/{highest \wt/} \def\hwm/{\hw/ \m/}
\def\gv/{generating vector} \def\BA/{Bethe-ansatz} \def\BAE/{\BA/ \eq/}
\def\evv/{$\E V$-valued} \def\wdn/{\wt/ decomposition}
\def\tp/{tensor product} \def\var/{variable} \def\resp/{respectively}
\def\fn/{function} \def\wf/{\wt/ \fn/} \def\opf/{one-point \fn/}
\def\Um/{$U_q$-\m/} \def\Uqm/{$\Uq$ \m/} \def\glm/{$\gl$ \m/}
\def\tri/{trigonometric} \def\rtl/{rational} \def\inl/{integral}
\def\J/{Jackson} \def\Ji/{\J/ \inl/} \def\Jir/{\Ji/ \rep/}
\def\ir/{\inl/ \rep/} \def\sol/{solution} \def\sev/{simple \egv/}
\def\KZ/{{\sl KZ\/}} \def\qKZ/{{\sl qKZ\/}}
\def\KZv/{Knizh\-nik-Zamo\-lod\-chi\-kov}
\def\aa/{associative algebra}

\def\q{{q\1}}
\def\va{v^\ast} \def\V{V^\ast} \def\G{G^\ast}
\def\dis{\{d_i(z;p)\}^s} \def\djs{\{d_j(z;p)\}^s} \def\dti{\tilde d}
\def\psit{\tilde\psi} \def\psis{\{\psi(z;p)\}^s}
\def\Psih{\hat\Psi} \def\Pch{\cdot\Psih(z;p)}
\def\oml{\om_{\la,V(1)\lc V(n)}(t,z)}
\def\oma{\om^\ast_{\la,\V(1)\lc\V(n)}(t,z)}

\def\gb{\bar\gm}
\def\gl{\frak{gl}_{N+1}} \def\Uq{U_q(\gl)}
\def\Yq{Y_q(\gl)} \def\yq{Y_q(\frak{gl}_N)}
\def\Nq{N_q(\gl)} \def\Vq{V_q(\gl)}
\def\Yg{Y(\gl)} \def\yg{Y(\frak{gl}_N)} \def\Ng{N(\gl)} \def\Vg{V(\gl)}

\def\Cn{\C^{N+1}} \def\Zn{\Z^N_{\ge 0}} \def\tu{T(u,1)} \def\Tu{\T(u,1)}
\def\En{\E(\Cn)}

\def\prn1{\prod^{N}_{i=1}} \def\prnl1{\prod^N_{l=1}}
\def\prlm1{\prod^{\la_i}_{j=1}} \def\prl{\prod^{\la_1}_{j=1}}
  \def\pu1j{(u_1(j))}
\def\ruu{R^{(i,j+1),(i,j)}}
\def\btij{\T_{I_0J_0}} \def\dij{_{I_0J_0}}
\def\pnp1{(N+1)} \def\ptu{(\tilde{u})}
\def\tpij{T^{(i,j)}} \def\tij{T^{(i,j)}}

\topmatter
\title
Jackson Integral Representations for Solutions to\\
the Quantized Knizhnik-Zamolodchikov Equation
\endtitle
\author
\vsk-.5>
V\. Tarasov$^\ast$ and A\. Varchenko$^{\ast\ast}$
\vsk.5>
\endauthor
\affil
${^\ast\,}$Physics Department, St\. Petersburg University\\
$^{\ast\ast\,}$Department of Mathematics, University of North Carolina
\endaffil
\address
\kern-\parindent \parindent 0pt
Physics Department\nl St\. Petersburg University\nl
St. Petersburg 198904, Russia \endaddress
\address Department of Mathematics\nl University of North Carolina\nl
Chapel Hill, NC 27599, USA \endaddress
\dedicatory
\vsk>
Dedicated to L. D. Faddeev on his 60$^{\text{th}}$ birthday
\enddedicatory
\thanks
{ }\vv-1>\nl
The first author was supported by Russian Foundation for Fundamental
Research.\nl
The second author was supported by NSF Grant DMS--9203929.\nl
\indent \llap{${^\ast}$}{\it E-mail: \sl tarasov\@onti.phys.lgu.spb.su}\nl
\indent \llap{$^{\ast\ast}$}{\it E-mail: \sl av\@math.unc.edu}
\endthanks

\abstract
The quantized \KZv/ \eq/s associated with the \tri/ \Rm/ of
the $\Uq$ type and the \rtl/ \Rm/ of the $\gl$ type are considered.
\Jir/s for \sol/s to these \eq/s are described. Asymptotic \sol/s to
a holonomic system of \deq/s are constructed. Relations between the \inl/
\rep/s and the \BA/ are indicated.
\endabstract

\endtopmatter

\document

\sect{} Introduction
\par
The \KZv/ \eq/ (\KZ/) is a holonomic system of \dfl/ \eq/s describing conformal
blocks in conformal field theory. This system has very rich mathematical
structures \cite{KZ, Koh, D2}. Most of these structures are revealed through
\ir/s for \sol/s to the \KZ/. Solutions are represented as multidimensional
hypergeometric \inl/s over cycles depending on parameters \cite{SV, V1}, and,
therefore, the \KZ/ is a special type of the Gauss-Manin connection.
\par
Recently, for several different reasons, the \KZv/ \eq/ was quantized \cite{S,
 FR, IJ}. The quantized \KZv/ \eq/ (\qKZ/) is a holonomic system of \deq/s. In
\cite{S} it is a system of \deq/s describing formfactors in quantum field
theory. In \cite{FR} it describes matrix elements of intertwining operators
for a quantum affine algebra. In \cite{IJ} it is a system of \eq/s for
correlation \fn/s of the six-vertex model. It is expected that the \qKZ/ has
remarkable mathematical properties as well as the \KZ/. The existence of \ir/s
for \sol/s to the \qKZ/ could be an important supporting argument for such
awaitings. Integral \rep/s for \sol/s might also be useful for applications
to the problems where the \qKZ/ were distilled.
\par
In this work, we describe \Ji/ \rep/s for \sol/s to the \qKZ/ associated with
$\gl$ and $\Uq$. The \Ji/ are \dc/ analogs of standard \inl/. We represent
\sol/s in terms of \dc/ multidimensional \inl/s of hypergeometric type (in
terms of multidimensional $q$-hypergeometric \fn/s), and therefore, we show
that quantized \KZv/ \eq/ are quantized Gauss-Manin connection. \Jir/s for
\sol/s to the \qKZ/ associated with $U_q(\frak{gl}_2)$ were described in
\cite{M1, M2, V2, R}.
\par
In \S1 we describe \sol/s to the \qKZ/ associated with $\Uq$, and in \S2 to
the \qKZ/ associated with
$\gl$. \S3 and \S4 contain proofs. In \S5 we construct asymptotic \sol/s to
a holonomic system of \deq/s, give a new formula for \BA/ eigenvectors in
a tensor product of $\gl$ or $\Uq$ modules and indicated connections of
the \Jir/s with the \BA/.
\par
This work was started when the authors visited RIMS in Kyoto. We thank
Professor T. Miwa and RIMS for their warm hospitality and stimulating
research atmosphere.

\head 1. Solutions to the \qKZ/. $\Uq$ Case \endhead

\sect(1.1) \qKZ/ for $\Uq$
\par
The quantum group $U_q=\Uq$ is the \aa/ generated by
$K^{\pm1}_1 \lc K^{\pm1}_{N+1}$ and $X^\pm_1 \lc X^\pm_N$
subject to the relations
$$
\NN2>
\align
[K_i,K_j]&=0\,, \qquad K_iK_i\1=K\1_iK_i=1\,,
\tag1.1.1\\
K_i X^\pm_i K\1_i&=q^{\pm1} X_i^\pm,\\
K_iX^\pm_{i-1} K\1_i&=q^{\mp 1} X_{i-1}^\pm\,,\\
K_iX^\pm_j K\1_i&=X^\pm_j \quad\ \text{for $j>i$ or $j<i-1$}\,,\\
[X^+_i,X^-_j]&=\dl_{ij}\frac{K_i K\1_{i+1}-K_{i+1}K\1_i}{q-\q}\,,\\
[X^\pm_i,X^\pm_j]&=0 \quad\ \text{for $|i-j|\ne1$}\,,\\
\nn1>
(X^\pm_i)^2 X^\pm_j-(q&+\q)X^\pm_i X^\pm_j X^\pm_i
+X^\pm_j(X^\pm_i)^2=0\,,\quad\ \text{for $|i-j|=1$}\,.
\endalign
$$
Here $q$ is a generic complex parameter.
\par
Let $\La=(\La_1\lc \La_{N+1})\in\Cn$. Let $V$ be a \Um/ with \hw/
$\La$, that is, $V$ is generated by one vector $v\in V$ such that
$$
X^+_iv=0\,,\quad K_iv=q^{\La_i}v
\tag1.1.2
$$
for all $i$. $V$ has the weight decomposition
\vvn-.2>
$$
\gather
V=\Plus_{\la\in\Zn}V_\la,
\tag1.1.3\\
\nn-9> \Text{where} \nn4>
V_\la=\{x\in V:K_i\,x=q^{\La_i+\la_{i-1}-\la_i}x\ \,\text{for all}\ i\}\,.
\tag1.1.4
\endgather
$$
 For any $\mu=(\mu_1\lc \mu_{N+1})\in\Cn$ introduce a linear operator
$$
\gather
L(\mu):V \to V
\tag1.1.5\\
\nn5> \Text{defined by} \nn-8>
L(\mu)\,x=q^{\sum^{N+1}_{i=1}\mu_i(\La_i-\la_i+\la_{i-1})}x
\tag1.1.6
\endgather
$$
for $x\in V_\la$.
\par
Let $\{V(m)\}$ be \Um/s with \hw/s $\{\La (m)\}$ and \gv/s $\{v_m\}$,
\resp/, $m=1\lc n$. The \wdn/ of \rep/s induces
the \wdn/ of the \tp/
$$
\align
&\kern37pt
V(1)\lox V(n)=\Plus_\la\ \bigl(V(1)\lox V(n)\bigr)_\la
\tagg"1.1.7"\\
\nn-4> \Text{where} \nn4>
&\bigl(V(1)\lox V(n)\bigr)_\la =
\Plus_{\la(1)+\ldots+\la(n)=\la} V(1)_{\la(1)}\lox V(n)_{\la(n)}\,.
\tagg"1.1.8"
\endalign
$$
\par
 For any $i,j$ there is the \tri/ \Rm/
$$
\NN3>
\align
R\0_{V(i)V(j)}(x): V(i)&\ox V(j) \to V(i)\ox V(j)
\tagg"1.1.9"\\
\Text{such that}
R\0_{V(i)V(j)}(x): v_i&\ox v_j \mapsto v_i\ox v_j\,,
\tagg"1.1.10"\\
R\0_{V(i)V(j)}(x)R\0_{V(i)V(k)}(xy) R\0_{V(j)V(k)}(y)
&=R\0_{V(j)V(k)}(y) R\0_{V(i)V(k)}(xy) R\0_{V(i)V(j)}(x)\,,
\endalign
$$
\cite{J, JKMO, C}, more precisely see in (3.2).
The \Rm/ preserves the weight decomposition. If
\vvn-.5>
$$
\gather
R\0_{V(i)V(j)}(x)=\sum_dR_i^{(d)}(x)\ox R_j^{(d)}(x)\,,\\
\nn2> \Text{then we let it act on $V(1)\lox V(n)$ by}
\sum_d^{\vp d}\,1\lox R_i^{(d)}(x) \lox R^{(d)}_j(x)\lox 1
\tag1.1.11
\vv-.5>
\endgather
$$
where $R_i^{(d)}(x)$ stands in the $i$-th factor and
$R_j^{(d)}(x)$ in the $j$-th factor.
\par
Denote by $L\0_{V(i)}(\mu)$ the linear operator on
$V(1)\lox V(n)$ acting as $L(\mu)$ on
the $i$-th factor and as the identity on all other factors.
\par
Let $p\in \C$ and $q=p^{-\nu}$
for some $\nu\in \C$. Let $Z_i$ denote the $p$-shift operator
$$
Z_i : \Psi (z_1\lc z_n) \mapsto \Psi(z_1\lc pz_i\lc z_n).
\tag1.1.12
$$
\flushpar
(1.1.13) {\it The quantized \KZv/ \eq/} (\qKZ/) for a
$V(1)\lox V(n)$-valued \fn/ $\Psi(z_1\lc z_n)$ is the system of \deq/s
$$
\align
Z_i\Psi&=R\0_{V(i)V(i-1)}\Bigl(\frac{pz_i}{z_{i-1}}\Bigr)\ldots
R\0_{V(i)V(1)} \Bigl(\frac{pz_i}{z_1}\Bigr)\x \\
&\quad\x q^{\al_i} L\0_{V(i)}(\mu)\,
R\1_{V(n)V(i)} \Bigl(\frac{z_n}{z_i}\Bigr) \ldots
R\1_{V(i+1)V(i)} \Bigl(\frac{z_{i+1}}{z_i}\Bigr)\,\Psi
\endalign
$$
for $i=1\lc n$. Here, $p,\al_1\lc \al_n\in \C$ and $\mu\in \Cn$ are
parameters of the \eq/, see \cite{FR}.
\par
In the next sections we give formulas for \sol/s to the $qKZ$. Namely, for any
$\la(1)\lc \la(n)\in \Z$, we construct
a $V(1)_{\la(1)}\lox V(n)_{\la(n)}$-valued
\vv.1> \fn/ \alh $w_{\la(1),V(1),\ldots,\la(n),V(n)}$
\vv.1> of certain \var/s $t_1,t_2,\ldots$ and
$z_1\lc z_n$ in such a way that the $V(1)\lox V(n)$-valued \fn/
$$
\Psi (z)=\sum_{\la(1)\lc \la(n)}
\int w_{\la(1),V(1)\lc \la(n),V(n)}(t,z)\,d_pt
$$
is a \sol/ to the \qKZ/. In this formula we use the notion of the \Ji/.

\sect (1.2) \J/ Integrals
\par
Discrete analogs of differentiation and integration are given by the formulas
$$
\gather
\frac{d_pf}{d_pt}(t)=\frac1t\ \frac{f(pt)-f(t)}{p-1}\,,
\tag1.2.1\\
\int^{\xi\8}_0 f(t)\,\frac{d_pt}{t}=(1-p)\sum^\8_{n=-\8} f(\xi p^n)\,.
\endgather
$$
The last sum is called the \Ji/ along a $p$-interval $[0,\xi\8]_p\,$.
There is a $p$-analog of the Stokes theorem:
$$
\int^{\xi\8}_0 t\,\frac{d_pf}{d_pt}\,\frac{d_pt}{t}=0\,.
\tag1.2.2
$$
\par
The multiple \Ji/ is defined similarly.\nl
$\Z^k$ acts on $\C^k$: for any $a=(a_1\lc a_k)\alb\in\Z^k$, set
$$
\align
T_a: \C^k &\to \C^k\,,
\tag1.2.3\\
(\xi_1\lc \xi_k) &\mapsto (\xi_1p^{a_1}\lc \xi_kp^{a_k})\,.\\
\nn3> \Text{$\Z^k$ acts on \fn/s on $\C^k$: for any $a\in\Z^k$, set} \nn4>
T_a : f(t_1\lc t_k) &\mapsto f(t_1p^{a_1}\lc t_kp^{a_k})\,.
\tag1.2.4
\endalign
$$
 For an arbitrary $\xi\in (\C^\ast)^k$, the $\Z^k$-orbit of $\xi$
is called a $k$-{\it dimensional} $p$-{\it cycle} and denoted by
$[0, \xi\8]_p$. The {\it \Ji/} of a \fn/ $f(t_1\lc t_k)$ over
a $p$-cycle $[0, \xi\8]_p$
$$
\int_{[0,\xi\8]_p} \<\< f(t_1\lc t_k)\,\Om\,,
\tag1.2.5
$$
for $\Om=(d_pt_1/t_1)\wedge\ldots\wedge(d_pt_k/t_k)$, is the sum
$$
(1-p)^k\< \sum_{-\8<a_1\lc a_k<\8} f(\xi_1p^{a_1}\lc \xi_kp^{a_k})
\tag1.2.6
$$
if it exists. For any $a\in\Z^k$, we have the Stokes formula
$$
\int_{[0,\xi\8]_p} \<\< T_a(f)\,\Om=\int_{[0,\xi\8]_p} \<\< f\,\Om\,.
\tag1.2.7
$$
\flushpar
(1.2.8) The \dif/ $T_a(f)\,\Om-f\,\Om$ will be called a \dc/ \dfl/.

\sect (1.3) One-Point Function of a Representation with Highest Weight
\par
Let $V$ be a \Um/ with \hw/ $\La$ and \gv/ $v$. For any
$\la=(\la_1\lc \la_N)\in \Zn$,
\vv.1>
we'll define a $V$-valued \rtl/ \fn/ $\eta_{\la,V}$ of \var/s
$\{u_i(j),y\}$, where $i=1\lc N$, $j=1\lc \la_i$.
\par
Let $R(x,y)\in \En\ox\En$ be defined by
$$
\multline
R(x,y)=\sum^{N+1}_{i=1} \al(x,y) E_{ii}\ox E_{ii}+\\
+\sum_{1\le i<j\le N+1}^{\vp N}
\bigl(\bt(x,y)\,(E_{ij}\ox E_{ij}+E_{ji}\ox E_{ji})
+\gm (x,y)E_{ij}\ox E_{ji}+\gb(x,y)E_{ji}\ox E_{ij} \bigr)
\endmultline
\tag1.3.1
$$
where $E_{ij}$ is the $(N+1)\x (N+1)$-matrix with
one nonzero entry $(E_{ij})_{ij}=1$, $\al(x,y)=xq-y\q$, $\bt (x,y)=x-y$,
$\gm (x,y)=x(q-\q)$, $\gb(x,y)=y(q-\q)$, see \cite{J} and references there.
\par
Let $T(u)$ be the $(N+1)\x (N+1)$-matrix with the following entries
$$
\gather
T_{ii}(u)=(uK_i-K\1_i)/(q-\q)\,,
\tag1.3.2\\
\nn1>
T_{ij}(u)=uF_{ji}K_j\,,\qquad T_{ji}(u)=K_j\1E_{ij}\,,
\tag1.3.3
\endgather
$$
where $i<j$ and $F_{ji}$, $E_{ij}$ are defined below:
$$
\NN1>
\gather
E_{i,i+1}=X^+_i, \quad F_{i+1,i}=X^-_i,
\tag1.3.4\\
\nn2> \Text{for $i=1\lc N$,} \nn2>
{\align
E_{i,i+p}&=E_{i,i+p-1} E_{i+p-1,i+p}-qE_{i+p-1,i+p}E_{i,i+p-1}\,,
\tagg"1.3.5"\\
 F_{i+p,i}&=F_{i+p,i+1} F_{i+1,i}-\q F_{i+1,i}F_{i+p,i+1}\,.
\endalign}
\endgather
$$
\par
Let $\la\in \Zn\,$,
\vv.1>
$k=\lsym+\la_N$. Consider the tensor product $\En^{\ox k}$.
Enumerate its factors by pairs (1,1),\,\(1,2)$\lc\,$(1,$\la_1$),\,%
\(2,1),\,\(2,2)$\lc\,(2,\la_2)\lc\,(N,1)\lc\,(N,\la_N)$.
This list defines the lexicographical order on the set of the pairs:
$(i,j)<(l,m)$ if $(i,j)$ stands before $(l,m)$.
\par
 For any $(i,j)$ denote by $T^{(i,j)}(u)$ the element
$$
1\lox T(u)\lox 1\in\En^{\ox k}
\tag1.3.6
$$
where $T(u)$ stands in the $(i,j)$-th place. Denote by
\vv.1>
$R^{(i,j),(l,m)}(x,y)\in\En^{\ox k}$ the element
which acts on $(\Cn)^{\ox k}$ as $R(x,y)$ on the $(i,j)$-th and
$(l,m)$-th factors and as identity on all other factors, see (1.1.11).
\par
Set
$$
\T(u,y)= \prod^N_{i=1}\ \prod^{\la_i}_{j=1}\ T^{(i,j)}(u_i(j)/y)
\x\prod_{(i,j)>(l ,m)}R^{(i,j),(l,m)}(u_i(j),u_l(m))\,.
\tag1.3.7
$$
Here the first product is taken in the lexicographical
order on pairs $(i,j)$ defined above. The second
product is also taken in the lexicographical order:
a factor $R^{(i,j),(l ,m)}$ stands on the right side of a factor
$R^{(i',j'),(l',m')}$ if $(i,j)<(i',j')$ or
$(i,j)=(i',j')$ and $(l,m)<(l',m')$.
\par
Let
$$
\align
\T(u,y)= \sum\Sb I=(i_1\lc i_k)\\J=(j_1\lc j_k) \endSb
\T_{IJ} (u,y)\, E_{i_1j_1}&\lox E_{i_kj_k}.
\tag1.3.8\\
\Text{In other words, $\T_{IJ}$ may be defined by}
\T(u,y) e_{j_1} \lox e_{j_k}= \sum_I^{\vp I}\T_{IJ}(u,y)\,
e_{i_1}&\lox e_{i_k}
\tag1.3.9
\endalign
$$
where $e_1 \lc e_{N+1} $ is the canonical basis in $\Cn$. Let
$$
\NN1>
\align
I_0&=(\und{1\lc 1}_{\tsize\la_1}, \und{2\lc 2}_{\tsize\la_2}\lc
\und{N\lc N}_{\tsize\la_N})\,,
\tagg"1.3.10"\\
J_0&=(j_1\lc j_k)\,,\quad j_a=i_a+1\,.
\endalign
$$
Set
$$
\gather
\eta_{\la,V}(u,y)
=\prod^N_{i=1}\biggl([\la_i]_q! \prod^{\la_i}_{j=1}
\Bigl(\left(q^{\La_{i+1}}
u_i(j)/y-q^{-\La_{i+1}} \right)
u_i(j)/y \Bigr)\x
\tag1.3.11 \\
\x\prod^{\la_i}_{j=2} \ \prod^{j-1}_{p=1}\al(u_i(j),u_i(p))
\x\prod^N_{m=i+1}\ \prod^{\la_i}_{j=1}\
\prod^{\la_m}_{p=1} \bt(u_m(p),u_i(j))\biggr)\1
\T_{I_0J_0}(u,y)\,v
\endgather
$$
where $[a]_q!=[1]_q [2]_q\ldots [a]_q$ and $[b]_q=(q^b-q^{-b})/(q-\q)$.
 For $\la=0$, set $\eta_{\la,V}=v$.
\par
$\eta_{\la,V}$ is a $V_\la$-valued \rtl/ \fn/ of $u,y$. We call
$\eta_{\la,V}$ the {\it \opf/}. For $i\in\{1\lc N\}$,
the \fn/ $\eta_{\la,V}$ is a symmetric \fn/ of
$u_i(1)\lc u_i(\la_i)$, see (3.6.2).
\flushpar
(1.3.12) {\it Examples of \opf/s for $U_q(\frak{gl}_3)$.}
$$
\alds
\align
&\la=(1,0):\ \eta_{\la,V}(u,y)=
\frac{y\,q^{\La_2}}{q^{\La_2}u_1(1)-q^{-\La_2}y}\,X^-_1\,v\\
&\la=(0,1):\ \eta_{\la,V}(u,y)=
\frac{y\,q^{\La_3}}{q^{\La_3}u_2(1)-q^{-\La_3}y}\,X^-_2\,v\\
&\la=(2,0):\ \eta_{\la,V}(u,y)=\frac{y^2 q^{2\La_2+1}}
{[2]_q (q^{\La_2}u_1(1)-q^{-\La_2}y)(q^{-\La_2}u_1(2)-q^{-\La_2}y)}
\,(X^-_1)^2\,v\\
&\la=(0,2):\ \eta_{\la,V}(u,y)=\frac{y^2 q^{2\La_3+1}}
{[2]_q (q^{\La_3}u_2(1)-q^{-\La_3}y)(q^{\La_3}u_2(2)-q^{-\La_3}y)}
\,(X^-_2)^2\,v\\
&\la=(1,1):\ \eta_{\la,V}(u,y)=\frac{y^2 q^{\La_3}}
{(u_2(1)-u_1(1))(q^{\La_3}u_2(1)-q^{-\La_3}y)}\,\x \\
&\hphantom{\la=(1,1):\ \eta_{\la,V}(u,y)=(u_2(1)-u_1}\biggl(
\frac{q^{\La_2}u_2(1)-q^{-\La_2}y}{q^{\La_2}u_1(1)-q^{-\La_2}y}
X^-_2X^-_1-\q X^-_1X^-_2\biggr)\,v\,.
\endalign
$$
\flushpar
(1.3.13) {\it Example of \opf/ for $\Uq$ and $\la=(0\lc \la_i\lc 0)$.}
\vv.1> \nl
Let $\la_i=k$, then
$$
\eta_{\la,V}(u,y)=\frac{y^kq^{k\La_{i+1}+k(k-1)/2}}
{[k]_q!\prod^k_{j=1}(q^{\La_{i+1}} u_k(j)-q^{-\La_{i+1}}y)}
\,(X^-_i)^k\,v\,,
$$
cf. \cite{M2, V2}.

\sect(1.4) Weight Function of a Tensor Product of
Representations with Highest Weight
\par
Let $\{V(m)\}$ be \Um/s with \hw/s $\{\La(m) \}$ and \gv/s $\{v_m\}$,
\resp/, $m=1\lc n$. For any $\la(1)\lc \la(n) \in \Zn$,
\vv.1>
we'll define a $V(1)_{\la(1)}\lox V(n)_{\la(n)}$-valued
\rtl/ \fn/ called the \wf/. The \wf/ will be defined in terms of \opf/s
$\eta_{\la(1),V(1)}\lc \eta_{\la(n),V(n)}$ introduced in (1.3). Let
$$
\NN1>
\align
\la&=\la(1)+\ldots+\la(n)\,,\tag1.4.1\\
\la&=(\la_1\lc \la_N)\,,\\
\la(m)&=(\la_1(m)\lc \la_N(m))\,,\qquad m=1\lc n\,.\\
\Text{Set}
l_i(m)&=\la_i (1)+\ldots+\la_i(m)\,,\qquad l_i(0)=0\,.
\tag1.4.2
\endalign
$$
We have $l_i(n)=\la_i$. Set
$$
\NN1>
\align
A(u,v) &=(uq-v\q)/(u-v)\,,\tag1.4.3 \\
B(u,v) &=(uq-v\q)/(u\q-vq)\,,\\
C_{i,m}(u,v) &=(uq^{\La_i(m)}-vq^{-\La_i(m)})/
(uq^{\La_{i+1}(m)}-vq^{-\La_{i+1}(m)})
\endalign
$$
where $\La_1(m)\lc \La_{N+1}(m)$ are coordinates of $\La(m)$.
\par
The \opf/ $\eta_{\la(m),V(m)}$ is a \fn/ of \var/s
$u_1(1)\lc u_1(\la_1(m))\lc u_N(1)\lc u_N(\la_N(m)),\alb\,y$, see (1.3).
Let $\si=(\si(1)\lc \si(N))$ where $\si(i)$ is an element of the symmetric
group $S_{\la_i}$. Define a \fn/ $\eta_{\la(m),V(m),\si}$ as the \fn/
$\eta_{\la(m),V(m)}$ where $y$ is replaced by $z_m$ and \var/s
$u_i(1)\lc u_i(\la_i(m))$ are replaced by the new \var/s
$\{t_i(\si_j (i))\}$, $j=l_i(m-1)+1\lc l_i(m)\,$ for $i=1\lc N$.
Introduce a $V(1)_{\la(1)}\lox V(n)_{\la(n)}$-valued \fn/ of \var/s
$t_1(1)\lc t_1(\la_1)\lc t_N(1)\lc t_N(\la_N),\,\alb z_1\lc z_n$
by the rule
$$
\alds
\gather
\hleft(1.4.4):{w_{\la(1),V(1)\lc \la(n),V(n);\si}=}
\tag1.4.4\\
= \prod^n_{j=2}\ \prod^{j-1}_{m=1}\ \prod^N_{i=1}
\ \prod^{l_i(j)}_{l=l_i(j-1)+1} C_{i,m}(t_i(\si_l(i)),z_m)\x
\prod^N_{i=1}\ \prod\Sb 1\le a<b\le\la_i\\ \si_a(i)>\si_b(i)\endSb
B(t_i(\si_a(i)),t_i(\si_b(i))) \x \\
{\align
\x \prod^n_{j=2} \ \prod^{j-1}_{m=1}\ \prod^N_{i=2}
\ \prod^{l_i(j)}_{l=l_i(j-1)+1} \ \prod^{l_i(m)}_{k=l_{i-1}(m-1)+1}
A&(t_i(\si_l(i)) ,t_{i-1}(\si_k(i-1))) \x \\
&\x \eta_{\la(1),V(1),\si}\lox \eta_{\la(n),V(n),\si}\,.
\endalign}
\endgather
$$
Define the {\it weight \fn/} by
$$
\align
&w_{\la(1),V(1)\lc \la(n),V(n)}(t,z)=
\sum_\si w_{\la(1),V(1)\lc \la(n),V(n);\si} (t,z)
\tagg"1.4.5" \\
\nn1> \Text{where the sum is taken over all
$\si=(\si(1)\lc \si(N))\in S_{\la_1}\lx S_{\la_N}$. Set}
&w_{\la,V(1)\lc V(n)} (t,z)=
\sum_{\la(1)+\ldots+\la(n)=\la}^{\vp\la}w_{\la(1),V(1)\lc \la(n),V(n)}(t,z)\,.
\tagg"1.4.6"
\endalign
$$
\Example
 For $n=2$, $\la(1)=(1,0\lc 0)$, $\la(2)=(0,1,0\lc 0)$, we have
$$
\NN2>
\gather
{\align
w_{\la(1),V(1),\la(2),V(2)}=
C_{2,1}(t_2(1),z_1) A(t_2(1),t_1(1))\x& \\
\x \eta_{\la(1),V(1)}(t_1(1),z_1)\ox& \eta_{\la(2),V(2)} (t_2(1),z_2)\,.
\endalign}\\
\nn3> \Text{For $n=2$, $\la(1)=(1,0\lc 0)$, $\la(2)=(1,0\lc 0)$, we have}
\nn3>
\hleft:{w_{\la(1),V(1),\la(2),V(2)}=C_{1,1}(t_1(2),z_1)\,
\eta_{\la(1),V(1)}(t_1(1),z_1)\ox \eta_{\la(2),V(2)} (t_1(2),z_2)+}\\
\nn1>
\hright:{+C_{1,1}(t_1(1),z_1) B(t_1(2),t_1(1))
\eta_{\la(1),V(1)} (t_1(2),z_1)\ox \eta_{\la(2),V(2)} (t_1(1),z_2)\,.}
\endgather
$$
\endexample
Now we'll define a \fn/ $\Phi$ of \var/s
$t_1(1)\lc t_1(\la_1)\lc t_N(1)\lc t_N(\la_N)$, $z_1\lc z_n$. Namely, set
$$
\alds
\NN2>
\align
(u)_\8 &=(u,p)_\8=\prod^{\8}_{m=0}(1-p^mu)\,.
\tagg"1.4.7"\\
\Text{Set}
\Phi_{t_i(j),z_m} &=
\frac{(q^{2\La_{i+1}(m)}t_i(j)/z_m,p)_\8}{(q^{2\La_i(m)}t_i(j)/z_m,p)_\8}\,
\biggl(\frac{t_i(j)}{z_m}\biggr)^{\<\nu(\La_i(m)-\La_{i+1}(m))}\,
\tagg"1.4.8" \\
\Phi_{t_i(a),t_i(b)} &=
\frac{(q^2 t_i(a)/t_i(b),p)_\8}{(q^{-2}t_i(a)/t_i(b),p)_\8}\,
\biggl(\frac{t_i(b)}{t_i(a)} \biggr)^{\< 2\nu},\\
\Phi_{t_i(a),t_{i+1}(b)} &=
\frac{(t_{i+1}(b)/t_i(a),p)_\8}{(q^2t_{i+1}(b)/t_i(a),p)_\8}\,
\biggl(\frac{t_{i+1}(b)}{t_i(a)} \biggr)^{\< \nu},
\endalign
$$
Let $\al_1\lc \al_n\in \C$ and $\mu=(\mu_1\lc \mu_{N+1})\in \Cn$. Set
$$
\multline
\Phi(t,z)=
\prod^n_{m=1}z_m^{-\nu\bigl(\al_m + \sum^{N+1}_{i=1}\mu_i\La_i(m)\bigr)}\x
\prod^{N}_{i=1}\ \prod^{\la_i}_{j=1}\ (t_i(j))^{-\nu(\mu_{i+1}-\mu_i)}\x \\
\x \prod^{n}_{m=1} \ \prod^{N}_{i=1}\ \prod^{\la_i}_{j=1}\ \Phi_{t_i(j),z_m}\x
\prod^{N-1}_{i=1}\ \prod^{\la_i}_{a=1}\ \prod^{\la_{i+1}}_{b=1}
\ \Phi_{t_i(a),t_{i+1}(b)}\x
\prod^{N}_{i=1}\ \prod_{1\le a<b\le \la_i}\<\Phi_{t_i(a),t_i(b)}\,.
\endmultline
\tag1.4.9
$$

\sect (1.5) \J/ Integral Representation of Solutions to the \qKZ/ for $\Uq$
\par
Let $\{V(m) \}$ be the \Um/s considered in (1.4). Let $\la\in \Zn$.
Set $k=\la_1\lsym+\la_N$. Consider $\C^k$ with coordinates
$t_1(1)\lc t_1(\la_1)\lc t_N(1)\lc t_N(\la_N)$.
 Fix a $k$-dimensional $p$-cycle $[0,\xi\8]_p$ in $\C^k$. Consider the
$V(1)\lox V(n)$-valued \fn/ $\Psi(z_1\lc z_n)$ defined by the formula
$$
\Psi(z)=\int_{[0,\xi\8]_p} \<\Phi(t,z)\,w_{\la,V(1)\lc V(n)}(t,z)\,\Om\,.
\tag1.5.1
$$
Here the \fn/s $w,\,\Phi$ are the \fn/s defined in (1.4.6) and (1.4.9),
\resp/. The \inl/ is the \Ji/ defined in (1.2).

\proclaim{(1.5.2) Theorem}
Assume that the \inl/ in {\rm (1.5.1)} exists. Then
the \fn/ $\Psi$ is a \sol/ to the \qKZ/ {\rm (1.1.13)}.
\endproclaim
\Remark{(1.5.3)}
More precisely, we prove that for any $i$
$$
\biggl(Z_i-R\0_{V(i),V(i-1)}\Bigl(\frac{pz_i}{z_{i-1}}\Bigr)\ldots
R\1_{V(i+1),V(i)}\Bigl(\frac{z_{i+1}}{z_i}\Bigr)\biggr)\,
\Phi(t,z)\,w_{\la,V(1)\lc V(n)}(t,z)\,\Om
$$
is a linear combination of \dc/ \dfl/s, cf\.\ (1.2.8).
In this work we will not discuss convergence of the \Ji/s.
\endremark
Theorem (1.5.2) is proved in \S4.

\head 2. Solutions to the \qKZ/, $\gl$ Case
\endhead

In this section we describe \sol/s to the \qKZ/ associated with $\gl$.
Constructions are completely parallel to the quantum group case described
in Section 1.

\sect (2.1) \qKZ/ for $\gl$
\par
$\gl$ is the Lie algebra generated by $k_1\lc k_{N+1}$,
$X^\pm_1\lc X^\pm_N$ subject to the relations
$$
\alignat2
[k_i,k_j]&=0\,,&&
\tag2.1.1\\
[k_i,X_i^\pm]&=\pm X_i^\pm, \qquad &&[k_i,X_{i-1}^\pm]=\mp X^\pm_{i-1}\,,\\
[k_i,X_j^\pm]&=0 \qquad &&\text{for $j>i$ or for $j<i-1$}\,,\\
[X^+_i,X^-_j]&=\dl_{ij}(k_i-k_{i+1})\,, \kern-20pt&&\\
[X_i^\pm,X^\pm_j]&=0 \qquad &&\text{for $|i-j|\ne 1$}\,,\\
[X_i^\pm,[X^\pm_i,X_j^\pm]] \kern-26pt &\kern26pt =0
\qquad &&\text{for $|i-j|=1$}\,.
\endalignat
$$
\par
Let $\La=(\La_1\lc \La_{N+1})\in \Cn$. Let $V$ be a \glm/ with \hw/ $\La$,
that is, $V$ is generated by a vector $v\in V$ such that
\vvn-.5>
$$
\NN2>
\gather
X^+_iv=0\,,\quad k_iv=\La_iv\,. \\
\Text{$V$ has the weight decomposition}
V=\bigoplus_{\la\in \Zn}V_\la
\endgather
$$
where
$V_\la=\{x\in V:k_ix=(\La_i+\la_{i-1}-\la_i)x$ for all $i\}$.
\par\nt
 For any $\mu=(\mu_1\lc \mu_{N+1})\in \Cn$ introduce a linear operator
$$
\gather
L(\mu): V\to V\,,
\tag2.1.2\\
x\mapsto \exp\Bigl(\,\tsum^{N+1}_{i=1} \mu_i(\La_i+\la_{i-1}-\la_i)\Bigr)\,x
\endgather
$$
for $x\in V_\la$.
\par
Let $\{V(m)\}$ be \glm/s with \hw/s $\{\La(m)\}$ and \gv/s $\{v_m\}$,
\resp/, $m=1\lc n$. For any $i,j$, there is the \rtl/ \Rm/
$$
\NN2>
\gather
{\align
R\0_{V(i)V(j)}(x): V(i)&\ox V(j) \to V(i)\ox V(j)
\tag2.1.3\\
\Text{such that}
R\0_{V(i)V(j)}(x) : v_i&\ox v_j \mapsto v_i \ox v_j\,,
\tag2.1.4
\endalign}\\
R\0_{V(i)V(j)}(x) R\0_{V(i)V(k)}(x+y) R\0_{V(j)V(k)}(y)=
R\0_{V(j)V(k)}(y) R\0_{V(i)V(k)}(x+y) R\0_{V(i)V(j)}(x)
\endgather
$$
\cite{KRS, JKMO, C}, more precisely, see in (3.7).
\par
Let $p\in \C$ and let $Z_i$ denote the $p$-shift operator
$$
Z_i: \Psi(z_1\lc z_n) \mapsto \Psi (z_1\lc z_i+p\lc z_n)\,.
\tag2.1.5
$$
\flushpar
(2.1.6) The quantized \KZv/ \eq/ for a $V(1)\lox V(n)$-valued
\fn/ $\Psi(z_1\lc z_n)$ is the system of \deq/s
$$
\multline
Z_i\Psi=R\0_{V(i)V(i-1)} (z_i-z_{i-1}+p)\ldots R\0_{V(i)V(1)}(z_i-z_1+p)\x \\
\x \exp(\al_i)\,L\0_{V(i)}(\mu)\,
R\1_{V(n)V(i)}(z_n-z_i)\ldots R\1_{V(i+1)V(i)}(z_{i+1}-z_i)\,\Psi
\endmultline
$$
for $i=1\lc n$. Here $p,\,\al_1\lc \al_n\in\C$ and $\mu\in\Cn$ are
parameters of the \eq/.
\par
In the next section, we give formulas for \sol/s to the $qKZ$.

\sect(2.2) Additive Version of \J/ Integrals
\par\nt
$\Z^k$ acts on $\C^k$: for any $a=(a_1\lc a_k)\in\alb\Z^k$, set
\vvn-.3>
$$
\align
Q_a: \C^k &\to \C^k\,,
\tag2.2.1\\
(\xi_1\lc \xi_k) &\mapsto (\xi_1+a_1p\lc \xi_k+a_kp)\,,\\
\nn2>
\Text{$\Z^k$ acts on \fn/s on $\C^k$: for any $a\in\Z^k$ set}
\nn3>
Q_a: f(t_1\lc t_k) &\mapsto f(t_1+a_1p\lc t_k+a_kp)\,.
\tag2.2.2
\endalign
$$
 For any $\xi\in \C^k$, the $\Z^k$-orbit of $\xi$ is called an
{\it additive $k$-dimensional $p$-cycle} and denoted by $[-\8,\xi\8]_p$.
The {\it additive \Ji/} of a \fn/ $f(t_1\lc t_k)$ over a $p$-cycle
$[-\8,\xi\8]_p$ is the sum
$$
\int_{[-\8,\xi\8]_p} \<\<\< f(t_1\lc t_k)\,d_pt_1\wedge \ldots\wedge d_pt_k=
p^k \<\< \sum_{-\8<a_1\lc a_k<\8} \< f(\xi_1+a_1p\lc \xi_k+a_kp)\,.
\tag2.2.3
$$
if it exists. For any $a\in \Z^k$, we have the Stokes formula
$$
\int_{[-\8,\xi\8]_p} \<\<\< Q_af\,d_pt= \int_{[-\8,\xi\8]_p} \<\<\< f\,d_pt\,.
\tag2.2.4
$$
\flushpar
{(2.2.5)} The \dif/ $Q(a)f\,d_pt-f\,d_pt$ will be called a \dc/ \dfl/.

\sect (2.3) One-Point Function of a Representation with Highest Weight
\par
Let $V$ be \glm/ with \hw/ $\La$ and \gv/ $v$.
Let $R(x,y)\in\E(\Cn)\ox \E(\Cn)$ be defined by (1.3.1) where
$\al(x,y)=x-y+1$, $\bt (x,y)=x-y$, $\gm (x,y)=\gb(x,y)=1$, \cite{Y}.
\par
Let $T(u)$ be the $(N+1)\x (N+1)$ matrix with the following entries:
$$
\gather
T_{ii}(u)=u+k_i\,,
\tag2.3.1\\
T_{ij}(u)=F_{ji}\,, \qquad T_{ji}(u)=E_{ij}\,,
\endgather
$$
where $i<j$ and $F_{ij}$, $E_{ij}$ are defined below:
\vvn-.3>
$$
\NN1>
\gather
E_{i,i+1}=X^+_i\,,\qquad F_{i+1,i}=X^-_i,
\tag2.3.2\\
\nn2> \Text{for $i=1\lc N$,} \nn2>
{\align
E_{i,i+p} &=[E_{i,i+p-1},E_{i+p-1,i+p}]\,,
\tag2.3.3 \\
 F_{i+p,i} &=[F_{i+p,i+1},F_{i+1,i}]\,.
\endalign}
\endgather
$$
\par
Let $\la\in \Zn,\ k=\la_1\lsym+ \la_N$. Set
$$
\T(u,y)=\prod^N_{i=1}\ \prod^{\la_i}_{j=1} T^{(i,j)}(u_i(j)-y)
\x \prod_{(i,j)<(l,m)} R^{(i,j),(l,m)} (u_i(j),u_l(m))\,,
\tag2.3.4
$$
where the products are taken in the same order as in (1.3.7). Set
$$
\multline
\eta_{\la,V}(u,y)=
\prod^N_{i=1}\Bigl(\la_i!\,\prod^{\la_i}_{j=1}(u_i(j)+\La_{i+1}-y)\x
\prod^{\la_i}_{j=2}\ \prod^{j-1}_{p=1}\al(u_i(j),u_i(p))\x \\
\x \prod^{N}_{m=i+1}\ \prod^{\la_i}_{j=1}\ \prod^{\la_m}_{p=1}
\bt (u_m(p),u_i(j))\Bigr)\1\T_{I_0J_0}(u,y)\,v\,.
\endmultline
\tag2.3.5
$$
 For $\la=0$, set $\eta_{\la,V}=v$.
\par
$\eta_{\la,V}$ is a \fn/ of $u,y$ with values in $V_\la$. For
$i\in\{1\lc N\}$, the \fn/ $\eta_{\la,V}$ is a symmetric \fn/ of
$u_i(1)\lc u_i(\la_i)$, cf\.\ (3.6.2).
$\eta_{\la,V}$ is called the {\it \opf/}.
\flushpar
{(2.3.6)} {\it Examples of \opf/s for $\frak{gl}_3$.}
$$
\NN1>
\alds
\align
\la=(1,0)&:
\ \eta_{\la,V}(u,y)=\frac{1}{u_1(1)+\La_2-y}\,X^-_1\,v\,. \\
\la=(0,1)&:
\ \eta_{\la,V}(u,y)=\frac{1}{u_2(1)+\La_3-y}\,X^-_2\,v\,. \\
\la=(2,0)&:
\ \eta_{\la,V}(u,y)=\frac{1/2}{(u_1(1)+\La_2-y)(u_1(2)+\La_2-y)}\,
(X^-_1)^2\,v\,. \\
\la=(0,2)&:
\ \eta_{\la,V}(u,y)=\frac{1/2}{(u_2(1)+\La_3-y)(u_2(2)+\La_3-y)}\,
(X^-_2)^2 \,v\,.\\
\la=(1,1)&:
\ \eta_{\la,V}(u,y)=\frac{1}{(u_2(1)-u_1(1))(u_2(1)+\La_3-y)}\,\x\\
&\hphantom{\eta_{\la,V}(u,y)=(u_2(1)-u_1(1))}
\biggl(\frac{u_2(1)+\La_2-y}{u_1(1)+\La_2-y}
X^-_2 X^-_1-X^-_1X^-_2 \biggr)\,v\,.
\endalign
$$
\flushpar
{(2.3.7)} {\it Example of a \opf/ for $\gl$ and
${\la}=(0\lc \la_i\lc 0)$.}
\vv.1>\nl
Let $\la_i=k$, then
\vvn-.2>
$$
\eta_{\la,V}(u,y)=\frac{1}{k!}\,\prod^k_{j=1}\,\frac{1}{u_k(j)+\La_{i+1}-y}
\,(X^-_i)^k\,v\,.
$$

\sect (2.4) Weight Function of a Tensor Product of
Representations
\par
Let $\{V(m)\}$ be \glm/ with \hw/s $\{\La(m)\}$ and \gv/s $\{v_m \}$,
\resp/, $m=1\lc n$. Set
\vvn-.5>
$$
\NN1>
\align
A(u,v)&=\frac{u-v+1}{u-v}\,,
\tag2.4.1 \\
B(u,v)&=\frac{u-v+1}{u-v-1}\,,\\
C_{i,m}(u,v)&=\frac{u-v+\La_i(m)}{u-v+\La_{i+1}(m)}\,.
\endalign
$$
\par
 For any $\la(1)\lc \la(n)\in\Zn$,
\vv.1>
define a \fn/ $w_{\la(1),V(1),\ldots,\la(n),V(n)}(t,z)$ by formula (1.4.5).
$w_{\la(1),V(1),\ldots,\la(n),V(n)}(t,z)$ is
a $V(1)_{\la(1)}\lox V(n)_{\la(n)}$-valued
\vv.1>
\fn/ of \var/s $\{t_i(j),z_m\}$, $i=1\lc N$, $j=1\lc \la_i(1)\lsym+\la_i(n)$,
$m=1\lc n$. Define the \wf/ $w_{\la,V(1)\lc V(n)}(t,z)$ by formula (1.4.6).
\par
Now we'll define a \fn/ $\Phi(t,z)$ which is an
additive analog of the \fn/ $\Phi$ defined in (1.4). Set
\vvn-.5>
$$
\alds
\NN2>
\align
\Phi_{t_i(j),z_m} &=
\frac{\Gm((t_i(j)-z_m+\La_i(m))/p)}{\Gm((t_i(j)-z_m+\La_{i+1}(m))/p)}\,,
\tag2.4.2 \\
\Phi_{t_i(a),t_i(b)} &=
\frac{\Gm((t_i(a)-t_i(b)-1)/p)}{\Gm((t_i(a)-t_i(b)+1)/p)}\,, \\
\Phi_{t_i(a),t_{i+1}(b)} &=
\frac{\Gm((t_{i+1}(b)-t_i(a)+1)/p)}{\Gm((t_{i+1}(b)-t_i(a))/p)}\,,
\endalign
$$
where $\Gm$ is the gamma-\fn/. Let $\al_1\lc \al_n\in \C$ and
$\mu=(\mu_1\lc \mu_{n+1})\in\Cn$. Set
\vvn-.5>
$$
\align
\quad \Phi(t,z)=& \prod^n_{m=1}\,
\exp\Bigl(z_m \bigl(\al_i+\tsum^{N+1}_{i=1}\mu_i\La_i(m)\bigr)/p \Bigr)
\,\x
\tag"\rlap{(2.4.3)}" \\
\x& \prod^{N}_{i=1}\ \prod^{\la_i}_{j=1}
\ \exp\big(t_i(j)(\mu_{i+1}-\mu_i)/p\big) \x
\prod^n_{m=1}\,\prod^{N}_{i=1}\,\prod^{\la_i}_{j=1}\,\Phi_{t_i(j),z_m}\,\x \\
\x& \prod^{N-1}_{i=1}\ \prod^{\la_i}_{a=1}\ \prod^{\la_{i+1}}_{b=1}
\Phi_{t_i(a),t_{i+1}}(b)\x
\prod^{N}_{i=1}\ \prod_{1\le a<b\le \la_i}\< \Phi_{t_i(a),t_i(b)}\,.
\endalign
$$

\sect (2.5) Integral Representation of Solutions to the \qKZ/ for $\gl$
\par
Let $\{V(m)\}$ be the \glm/s considered in (2.4). Let $\la\in \Zn$. Set
$k=\la_1\lsym+ \la_N$. Consider $\C^k$ with coordinates $\{t_i(j),z_m\}$.
 Fix an additive $k$-dimensional $p$-cycle $[-\8,\xi\8]_p$ in $\C^k$.
Consider the $V(1)\lox V(n)$-valued \fn/ $\Psi(z_1\lc z_n)$ defined by
the formula
$$
\Psi(z)=\int_{[-\8,\xi\8]_p}\<\< \Phi(t,z)\,w_{\la,V(1)\lc V(n)}(t,z)
\,d_pt_1(1)\wedge\ldots\wedge d_pt_k(\la_k)\,.
\tag2.5.1
$$
Here \fn/s $\Phi,w$ are the \fn/s defined in (2.4). The \inl/ is
the additive \Ji/ defined in (2.2).
\proclaim{(2.5.2) Theorem}
Assume that the \inl/ in {\rm (2.5.1)} exists.
Then the \fn/ $\Psi$ is a \sol/ to the \qKZ/ {\rm (2.1.6)}.
\endproclaim
\Remark{(2.5.3)}
More precisely, we prove that for any $i$
$$
\bigl(Z_i-R\0_{V(i),V(i-1)} (z_i-z_{i-1}+p)\ldots
R\1_{V(i-1),V(i)} (z_{i+1}-z_i)\bigr)
\,\Phi(t,z)\,w_{\la,V(1)\lc V(n)}(t,z)\,d_pt
$$
is a linear combination of \dc/ \dfl/s.
\endremark
The proof of Theorem (2.5.2) is completely analogous to
the proof of Theorem (1.5.2).
\par
Consider $V(1)\lox V(n)$ as a \glm/.
\proclaim{(2.5.4) Lemma}
Assume that \inl/ in {\rm (2.5.1)} exist. Then for any
$i=1\lc N$ the equality $\mu_i=\mu_{i+1}$ implies that
$X_i^-\Psi(z)$ is a \sol/ to \qKZ/ {\rm (2.1.6)} and $X_i^+\Psi(z)=0$.
\endproclaim
\Proof.
It is obvious that $X_i^-$ commutes with the operator in \rhs/ of \qKZ/
(2.1.6) if $\mu_i=\mu_{i+1}$. Then the first equality simply follows from
Theorem (2.5.2). The proof of the second one will be given elsewhere.
\endproof
\Remark{(2.5.5)}
If $\mu_1=\ldots=\mu_{N+1}$ then $\Psi(z)$ is a \hw/ vector of
an irreducible $\gl$ submodule in $V(1)\lox V(n)$.
\endremark

\head 3. $R$-Matrix Property of Weight Functions
\endhead

\sect (3.1) Weight Functions of the $\Uq$ Case
\par
In this section we prove the \Rm/ property for the \wf/s
\alh $w_{\la,V(1)\lc V(n)}(t,z)$ introduced in (1.4.6).
\par
Let $V(1)\lc V(n)$ be $\Uq$ \hwm/s, $R\0_{V(i),V(j)}$ the \tri/ \Rm/ (1.1.10),
(1.1.11), (3.2.14). For $i=1\lc N$ let
$$
\align
P_i: V(1)\lox V(i+1) \ox V(i)\lox V(n) &\to V(1)\lox V(n)\,,\\
\nn1>
x_1\lox x_{i+1}\ox x_i \lox x_n &\mapsto x_1\lox x_n
\endalign
$$
be the \perm/ of factors and $\tilde{z}_i=\(z_1\lc z_{i+1},z_i\lc z_n)$.
\proclaim{(3.1.1) Theorem}
We have
$$
R\0_{V(i),V(i+1)}(z_i/z_{i+1})\,w_{\la,V(1)\lc V(n)}(t,z)=
P_i\,w_{\la,V(1)\lc V(i+1),V(i)\lc V(n)}(t,\tilde{z}_i)\,.
$$
\endproclaim
The theorem will be proved in (3.6.9).

\sect (3.2) Quantum Yangian
\par
The \Rm/ $R(x,y)$ defined by (1.3.1) satisfies the \YB/
$$
\align
R^{(1,2)}(x_1,x_2) &R^{(1,3)}(x_1,x_3) R^{(2,3)}(x_2,x_3)=
\tagg"3.2.1"\\
\nn1>
\x &R^{(2,3)}(x_2,x_3) R^{(1,3)}(x_1,x_3) R^{(1,2)}(x_1,x_2)\,.
\endalign
$$
\par
The {\it quantum Yangian} $\Yq$ is the \aa/ generated by
elements $T^s_{ij}$, $\,i,j\in \{1\lc N+1\}$, $s=0,1\ldots$ subject to
the relations
\roster
\item"(3.2.2)" $\,T^0_{ij}=0\,$ for $i>j\,$,
\vv.1>
\item"(3.2.3)" $\,T^0_{ii}\,$ are invertible,
\vv.1>
\item"(3.2.4)"
$\,R(x,y)\,T^{(1)}(x)\,T^{(2)}(y)=T^{(2)}(y)\,T^{(1)}(x)\,R(x,y)$
\endroster
where $T^{(1)}(x)=T(x)\ox 1$, $T^{(2)}(y)=1\ox T(y)$, and $T(x)$ is
the $(N+1)\x (N+1)$-matrix with entries
$$
T_{ij}(x)=\sum^\8_{s=0} T^s_{ij}\,x^{1-s}\,.
$$
The quantum Yangian is isomorphic to the Borel subalgebra of
$U_q(\widehat{\frak{gl}}_{N+1})$ \cite{RS, DF}.
It is a \tri/ $R$-algebra of infinite level in the sense of \cite{C}.
\par
The generators $T^s_{ij}$ for $i,j=2\lc N+1$ obey
exactly the same relations as the generators of $\yq$. Denote by
$$
\align
\dl: \yq &\to \Yq \,,
\tag3.2.5\\
T^s_{ij} &\mapsto T^s_{i+1,j+1}\,,\\
\nn4>
\Text{the corresponding embedding. Relations (3.2.4) imply that the map}
\nn4>
\theta_y: \Yq &\to \Yq[y,y\1]\,,
\tag3.2.6\\
T^s_{ij} &\mapsto y^{s+j-i-1} T^s_{ij}\,,\\
\nn4>
\Text{is an embedding of algebras. Moreover, by virtue of (3.2.1) we have
a homomorphism}
\nn4>
\rho_y: \Yq &\to \En[y]\,,
\tag3.2.7\\
T(x) &\mapsto R(x,y)\,.\\
\nn5>
\Text{\endgraf The quantum Yangian $\Yq$ admits a natural coproduct}
\nn5>
\Dl: T_{ij}(x) &\mapsto x^{-1}\sum^{N+1}_{k=1} T_{ik}(x)\ox T_{kj}(x)\,.
\tag3.2.8
\endalign
$$
Let $\Dl^{(m)}$ be the $m$-iterated coproduct $(\Dl^{(0)}=\id$,
$\Dl^{(1)}=\Dl)$. Consider the natural action of the symmetric group
$S_{m+1}$ on $\Yq^{\ox(m+1)}$:
$$
\si(X_1\lox X_{m+1})=X_{\si_1}\lox X_{\si_{m+1}}.
$$
Set $\Dl^{(m),\si}=\si\o\Dl^{(m)}$.
\par
Let $\Nq$ be the left ideal in $\Yq$ generated by all $T^s_{ij}$ for $i>j$.
Set
$$
\Vq= \Yq \lower2pt\hbox{$\big/$} \lower4pt\hbox{$\Nq$}\,.
\tag3.2.9
$$
We will use the same letter for both an element of $\Yq$ and its projection
in $\Vq$. The symbol $\approx$ will be used if an equality takes place in
$\Vq$. One can check that $\Nq$ is a two-sided coideal in $\Yq$. Hence, the
coproduct $\Dl$ induces a coproduct
$$
\Vq \to \Vq\ox \Vq\,.
$$
\par
The embedding $\dl$, the map $\rho_y$, and the coproduct $\Dl$ have
the properties:
$$
\align
(\dl\ox\dl)\o\Dl(X)&\approx(\Dl\o\dl)(X)\,,
\tag3.2.10\\
\rho_y(X)&=(\rho_y\o\dl)(X)\vst{{\C^N}}\,,
\endalign
$$
for any $X\in\yq$, here $\C^N$ is supposed to be the span of the last $N$
canonical basic vectors in $\Cn$, $\rho_y$, $\dl$ and $\Dl$ in \lhs/s
are related to $\yq$, and $\rho_y$, $\dl$ and $\Dl$ in \rhs/s to $\Yq$.
\par
There exists a homomorphism
$$
\NN4>
\align
\phi: \Yq &\to \Uq
\tag3.2.11\\
\Text{defined by formulas (1.3.2) and (1.3.3). Set}
\phi_y=\phi\o\theta_y: \Yq &\to \Uq[y,y\1]\,.
\tag3.2.12
\endalign
$$
\par
Let $V_1$ and $V_2$ be $\Uq$ \hwm/s with \gv/s $v_1$ and $v_2$, \resp/.
There exists a linear map
$$
\NN2>
\gather
R\0_{V_1,V_2}(x): V_1\ox V_2\to V_1\ox V_2
\tag3.2.13\\
\Text{such that}
{\align
R\0_{V_1,V_2}(x/y)\,(\phi_x\ox\phi_y)\o\Dl(X) &=
(\phi_x\ox\phi_y)\o\Dl'(X)\,R\0_{V_1,V_2}(x/y)
\tagg"3.2.14"\\
\nn2> \Text{for any $X\in\Yq\,$ and} \nn2>
R\0_{V_1,V_2}(x)\,v_1\ox v_2 &=v_1\ox v_2\,,
\tagg"3.2.15"
\endalign}
\endgather
$$
see \cite{C}. Here $\Dl'=P\o\Dl$ where $P$ is the \perm/ of factors.
Moreover, $R\0_{V_1,V_2}(x)$ satisfies the YB \eq/ (1.1.10) and
$$
R\1_{V_2,V_1} (x\1)=P\0_{V_1,V_2}R\0_{V_1,V_2}(x)P\0_{V_2,V_1}
\tag3.2.16
$$
where $P\0_{V_i,V_j}: V_i\ox V_j\to V_j\ox V_i$ is the \perm/.
\vsk.1>
{\it This \Rm/ is used in the \qKZ/} (1.1.13).
\vsk.1>
\Remark{(3.2.17)}
Properties (3.2.14) and (3.2.15) uniquely define
the \Rm/ $R\0_{V_1,V_2}(x)$ for irreducible modules $V_1$ and $V_2$.
\endremark

\sect (3.3) Function $F_\la(u)$
\par
 For $\la\in \Zn$, define $\T(u,1)$ and $\T_{I_0J_0}(u,1)$ by formulas
(1.3.7)--\(1.3.10).
\proclaim{(3.3.1) Lemma}
There exists a unique formal Laurent series
$F_\la(u)$ in \var/s $\{u_i(j)\}$ with a finite number of positive powers
in each \var/, such that
$$
\T_{I_0J_0}(u,1)=F_\la(u)\times
\prod^{N}_{i=1}\ \prod_{j>m}\,\al(u_i(j),u_i(m))\,.
$$
\endproclaim
\Proof.
Using (3.2.1), we may rewrite (1.3.7) as follows:
$$
\T(u,1)=\prod^{N}_{i=1}\ \prod^{\la_i}_{j=1} \,T^{(i,j)}(u_i(j))
\x \Cal R(u)\x \prod^{N}_{i=1}\ \prod_{j>m}\,R^{(i,j),(i,m)}(u_i(j),u_i(m))\,,
\tag3.3.2
$$
where $\Cal R(u)$ is a polynomial matrix in $\{u_i(j)\}$
and the last product is in the lexicographical order.
 For example, for $\la=(2,2,0\lc 0)$, we have
$$
\NN1>
\align
&R^{(2,2),(2,1)}(u_2(2),u_2(1)) R^{(2,2),(1,2)}(u_2(2),u_1(2))
R^{(2,2),(1,1)}(u_2(2),u_1(1))\x \\
&R^{(2,1),(1,2)}(u_2(1),u_1(2)) R^{(2,1),(1,1)}(u_2(1),u_1(1))
R^{(1,2),(1,1)}(u_1(2),u_1(1))=\\
&=\bigl(R^{(2,1),(1,2)}(u_2(1),u_1(2)) R^{(2,2),(1,2)}(u_2(2),u_1(2))
R^{(2,1),(1,1)}(u_2(1),u_1(1))\x\\
&\x R^{(2,2),(1,1)}(u_2(2),u_1(1))\bigr)\cdot
R^{(2,2),(2,1)}(u_2(2),u_2(1)) R^{(1,2),(1,1)}(u_1(2),u_1(1))\,.
\endalign
$$
Let $J_0=(j_1\lc j_k)$, then
$$
\align
\Tu\,e_{j_1} \lox\,e_{j_k}=\prn1\ \prlm1\ T^{(i,j)}(u_i(j))\x \Cal R(u)
\,e_{j_1}\lox e_{j_k}\x&
\tagg"3.3.3"\\
\x \prn1\ \prod_{j>m}\,\al(u_i(j),u_i(m))& \,,
\endalign
$$
since $R(x,y)\,e_j\ox e_j=\al(x,y)\,e_j\ox e_j$. This proves the lemma.
\endproof
\proclaim{(3.3.4) Theorem}
 For any $i\in\{1\lc N\}$, the \fn/ $F_\la(u)$ is
a symmetric \fn/ of $u_i(1)\lc u_i(\la_i)$.
\endproclaim
\Proof.
 Fix $j\in \{1\lc \la_i-1\}$. Using (3.2.1) we may write
$$
\Tu=\Cal T(u) R^{(i,j+1),(i,j)}(u_i(j+1),u_i(j))
\tag3.3.4
$$
where $\Cal T(u)$ is $\prnl1\,\prod^{\la_l}_{m=1}\,T^{l ,m}(u_l(m))$
multiplied by some product of \Rms/. Using (3.2.1) and (3.2.4), we may write
$$
\Tu=\ruu(u_i(j+1),u_i(j))\,\overline{\Cal T}(u)
\tag3.3.5
$$
where $\overline{\Cal T}(u)$ is a suitable matrix.
\par
Let $P$ be the \perm/ in $\Cn\ox\Cn$: $P(x\ox y)=y\ox x$. Let
$\Cal T(\bar{u})$ be the \fn/ obtained from the \fn/ $\Cal T(u)$ by
interchanging $u_i(j)$ and $u_i(j+1)$.
\proclaim{(3.3.6) Lemma}
$\,\overline{\Cal T}(u)=P^{(i,j+1),(i,j)}\Cal T(\bar{u}) P^{(i,j+1),(i,j)}\,$.
\endproclaim
\Proof.
 For example, for $\la=(1,2,1)$ and $i=2$, $j=1$, we have
$$
\alds
\align
\tu &=T^{(1,1)}(u_1(1)) T^{(2,1)}(u_2(1)) T^{(2,2)}(u_2(2))
T^{(3,1)}(u_3(1)) \x \\
&\quad\x R^{(3,1),(2,2)}(u_3(1),u_2(2)) R^{(3,1),(2,1)}(u_3(1),u_2(1))
R^{(3,1),(1,1)}(u_3(1),u_1(1)) \x \\
&\quad\x R^{(2,2),(2,1)}(u_2(2),u_2(1)) R^{(2,2),(1,1)}(u_2(2),u_1(1))
R^{(2,1),(1,1)}(u_2(1),u_1(1))\,,\\
\nn2>
\Cal T(u)&=T^{(1,1)}(u_1(1)) T^{(2,1)}(u_2(1)) T^{(2,2)}(u_2(2))
T^{(3,1)}(u_3(1)) \x \\
&\quad\x R^{(3,1),(2,2)}(u_3(1),u_2(2)) R^{(3,1),(2,1)}(u_3(1),u_2(1))
R^{(3,1),(1,1)}(u_3(1),u_1(1)) \x \\
&\quad\x R^{(2,1),(1,1)}(u_2(1),u_1(1)) R^{(2,2),(1,1)}(u_2(2),u_1(1))\,.
\endalign
$$
The general proof can be done easily by induction on $\la_1\lc \la_N$.
\endproof
By (3.3.4) and (3.3.5), we have
$$
\btij(u,1)=\al(u_i(j+1),u_i(j))\,\Cal T\dij(u)=
\al(u_i(j+1),u_i(j))\,\Cal T\dij(\bar{u})\,,
$$
hence $F_\la(u)$ is symmetric with respect to the \perm/ of $u_i(j)$
and $u_i(j+1)$. The theorem is proved.
\endproof

\sect (3.4) Inductive Definition of $F_\la(u)$
\par
We give an inductive definition for $F_\la(u)$ which comes from the nested
\BA/ \cite{KR}.
\par
Consider the $\pnp1\x\pnp1$-matrix $T(x)$ as a $2\x2$ block matrix
$$
T(x)=\pmatrix T_{11}(x) &K(x)\\ \\L(x) &M(x) \endpmatrix
$$
where $K(x)$ is a row of length $N$, $L(x)$ is a column of length $N$,
and $M(x)$ is an $N\x N$-matrix. We will consider $K(x)$ as a linear map
$$
K(x): \C^N \to x\Yq [[x\1]]\,.
$$
As usual set $K^{(i)}(x)=1\lox K(x)\lox 1$, where $K(x)$ stands in the
$i$-th place. The matrix $M(x)$ is the image of the canonical
generators of $\yq$ under the embedding $\dl$, see (3.2.5).
\par
 For $\la\in \Zn$, define $\tilde{\la}\in \Z^{N-1}$, such that
$\tilde{\la}_i=\la_{i+1}$, and set $\tilde{u}_i(j)=u_{i+1}(j)$. Consider
$F_{\tilde{\la}}(\tilde{u})$ defined in $\yq$. Set
$$
\widetilde{F}_\la(u)=
\bigl(\dl\ox\rho_{{u_1(1)}}\lox\rho_{{u_1(\la_1)}}\bigr)\o
\Dl^{(\la_1),\si}\bigl(F_{\tilde{\la}}(\tilde{u})\bigr)
\tag3.4.1
$$
where $\si(1,2\lc \la_1+1)=(1,\la_1+1\lc 2)$. The coproduct $\Dl$, (3.2.8),
and the map $\rho$, (3.2.7), are taken for the smaller quantum Yangian $\yq$.
$\widetilde{F}_\la(u)$ is a matrix acting in $(\C^N)^{\ox\la_1}$ with
$\Yq$-valued entries.
\par
Let $e=(1,0\lc 0)^t$ be the first basic vector in $\C^N$.
\proclaim{(3.4.2) Theorem}
$$
 F_\la(u)\approx K^{(1)}(u_1(1))\ldots K^{(\la_1)}(u_1(\la_1))
\bigl(\widetilde{F}_\la(u)\,e\lox e\bigr)\x
\prod_{i=2}^N\ \prod_{j=1}^{\la_i}\,u^{\la_1}_i(j)\,.
$$
\endproclaim
\Proof.
Rewrite $\T(u,1)$ as follows:
$$
\alds
\multline
\T(u,1)=\prod^{\la_1}_{j=1} T^{(1,j)}(u_1(j)) \x
\prod^N_{i=2}\ \prlm1 \,T^{(i,j)}(u_i(j)) \x \\
\aligned
\x \prod\Sb(i,j)>(l,m)\\ l\ge 2\endSb\ R^{(i,j),(l ,m)}(u_i(j),u_l(m)) \x
\prod\Sb(i,j),m\\ i\ge 2\endSb\ R^{(i,j),(1,m)}(u_i(j),u_1(m)) \x& \\
\x \prod_{j>m}\ R^{(1,j),(1,m)}(u_1(j),u_1(m))&\,.
\endaligned
\endmultline
\tag3.4.3
$$
Here we use the commutativity of \Rms/ without
common superscripts. Applying (3.2.1), we may rewrite (3.4.3) as
$$
\alds
\multline
\T(u,1)=\prod^{\la_1}_{j=1} T^{(1,j)}(u_1(j)) \x \\
\aligned
\x \prod^N_{i=2}\ \prlm1\,\Bigl(T^{(i,j)}(u_i(j))
R^{(i,j),(1,\la_1)}(u_i(j),u_1(\la_1))\ldots
R^{(i,j),(1,1)}(u_i(j),u_1(1))\Bigr) \x& \\
\x \prod\Sb(i,j)>(l ,m)\\ l\ge 2 \endSb\ R^{(i,j),(l ,m)}(u_i(j),u_l(m)) \x
\prod_{j>m}\ R^{(1,j),(1,m)}(u_1(j),u_1(m))&\,.
\endaligned
\endmultline
\tag3.4.4
$$
 For example, for $\la=(1,3)$ we have
$$
\alds
\align
&R^{(2,3),(2,2)}(u_2(3),u_2(2)) R^{(2,3),(2,1)}(u_2(3),u_2(1))
R^{(2,3),(1,1)}(u_2(3),u_1(1)) \x \\
&\x R^{(2,2),(2,1)}(u_2(2),u_2(1)) R^{(2,2),(1,1)}(u_2(2),u_1(1))
R^{(2,1),(1,1)}(u_2(1),u_1(1))=\\
&=(R^{(2,3),(2,2)} R^{(2,3),(2,1)} R^{(2,2),(2,1)})(R^{(2,3),(1,1)}
R^{(2,2),(1,1)} R^{(2,1),(1,1)})=\\
&=R^{(2,1),(1,1)} R^{(2,2),(1,1)} R^{(2,3),(1,1)})
(R^{(2,3),(2,2)} R^{(2,3),(2,1)} R^{(2,2),(2,1)})\,.
\endalign
$$
\par
By the definition of $\rho_y$ in (3.2.7), we rewrite (3.4.4) as
$$
\alds
\multline
\Tu=\prlm1 \ T^{(1,j)}\pu1j \x \Bigl(\id\ox\rho_{u_1(1)}^{(1,1)}\lox
\rho_{u_1(\la_1)}^{(1,\la_1)}\Bigr)\,\o \\
\aligned \Dl^{(\la_1),\si}
\biggl(\,\prod^N_{i=2}\ \prlm1\ T^{(i,j)}(u_i(j)) \x
\prod\Sb(i,j)>(l ,m)\\ l\ge 2 \endSb
\ R^{(i,j),(l ,m)}(u_i(j),u_l(m))\biggr) \x& \\
\x \prod_{j>m} R^{(1,j),(1,m)}(u_1(j),u_1(m))\x
\prod_{i=2}^N\ \prod_{j=1}^{\la_i}\,u^{\la_1}_i(j)&\,,
\endaligned
\endmultline
\tag3.4.5
$$
where the superscript for $\rho^{(1,j)}_{u_1(j)}$ indicates the target space.
Introduce
$$
I_1=(\und{1\lc 1}_{\tsize \la_1})\quad \text{and}\quad
I_2=(\und{2\lc 2}_{\tsize \la_2}\lc \und{N\lc N}_{\tsize \la_N})
$$
so that $I_0=(I_1,I_2)$, and similarly for $J_0=(J_1,J_2)$.
Then, according to (3.4.5), we have
$$
\multline
\btij(u,1)=\biggl[\ \prlm1 \ T^{(1,j)}\pu1j \x
\left(\id\ox\rho^{(1,1)}_{u_1(1)} \lox
\rho^{(1,\la_1)}_{u_1(\la_1)}\right)\,\o \\
\aligned
\Dl^{(\la_1),\si} \biggl(\biggl[\ \prod^N_{i=2}\ \prlm1 T^{(i,j)}(u_i(j)) \x
\prod\Sb{(i,j)>(l ,m)}\\ l\ge 2 \endSb
R^{(i,j),(l ,m)}(u_i(j),u_l (m))\biggr]_{I_2J_2}\biggr)\biggr]_{I_1J_1}\x&\\
\x \prod_{j>m}\al (u_1(j),u_1(m))\x
\prod_{i=2}^N\ \prod_{j=1}^{\la_i}\,u^{\la_1}_i(j)\,.
\endaligned
\endmultline
\tag3.4.6
$$
Due to the explicit formula for the \Rm/ in (1.3.1),
$$
\biggl[\ \prod^N_{i=2}\ \prlm1 \tpij (u_i(j)) \x
\prod\Sb(i,j)>(l ,m)\\ l\ge 2 \endSb
R^{(i,j),(l ,m)}(u_i(j),u_l(m))\biggr]_{I_2J_2}\in \dl(\yq)
\tag3.4.7
$$
and only the part of the \Rm/ corresponding to $\dl(\yq)$ is employed in
this entry. Namely, this entry is equal to
$$
\align
&\quad\dl\biggl(\biggl[\ \prod^{N-1}_{i=1}\ \prod^{\tilde{\la}_i}_{j=1}
\tij(\tilde{u}_i(j)) \x
\prod\Sb(i,j)>(l,m)\\ l\ge 2\endSb
R^{(i,j),(l,m)}(\tilde{u}_i(j),\tilde{u}_l(m))\biggr]_{I_2,J_2}\biggr)=
\tagg"3.4.8"\\
&\quad=\dl(\T_{I_2,J_2}(\tilde{u},1))
\endalign
$$
where $\tij$, $R^{(i,j),(l,m)}$ and $\T_{I_2,J_2}$ are now related to $\yq$.
Now the theorem follows from (3.4.6)--\(3.4.8), (3.4.1) and (3.2.10):
$$
\multline
 F_\la(u)\x
\Bigl(\prod_{i=2}^N\ \prod_{j=1}^{\la_i}\,u^{\la_1}_i(j)\Bigr)^{-1}=\\
=\biggl[\ \prl\tij\pu1j \x
\left(\id\ox\rho^{(1,1)}_{u_1(1)}\lox\rho^{(1,\la_1)}_{u_1(\la_1)}\right)\o
\Dl^{(\la_1),\si}\o
\dl\left(F_{\tilde{\la}}\ptu \right)\biggr]_{I_1J_1}\approx \\
\ald
\aligned
\approx \sum_I \Bigl[\ \prl T^{(1,j)}\pu1j \Bigr]_{I_1I}\! \x
\Bigl[\,\bigl(\dl\ox\rho_{u_1(1)}\lox
\rho_{u_1(\la_1)} \bigr)\o \Dl^{(\la_1),\si}
\bigl(F_{\tilde{\la}}(\tilde{u})\bigr)\Bigr]_{IJ_1}=&\\
=K^{(1)}(u_1(1)) \ldots K^{(\la_1)}(u_1(\la_1))
\bigl(\widetilde{F}_\la(u)\,e\lox e\bigr)&\,.
\endaligned
\endmultline
\tag3.4.9
$$
\endproof

\sect (3.5) Comultiplication Properties of $F_\la(u)$
\par
\proclaim{(3.5.1) Theorem}
$$
\alds
\multline
\Dl F_{\la(u)} \approx \sum_{\Gm (1)\cup\Gm (2)}
\prn1\ \prod\Sb j\in \Gm_i(1)\\ l \in\Gm_i(2) \endSb
A (u_i(j),u_i(l)) \x \prod^{N-1}_{i=1}
\prod\Sb j\in\Gm_i(1)\\ l\in\Gm_{i+1}(2)\endSb\< A (u_{i+1}(l),u_i(j))\x \\
\aligned
\x \prod^{N-1}_{i=1}\ \prod^{N}_{m=i+1}
\prod \Sb j\in \Gm_i(1)\\l \in \Gm_m(2) \endSb \bt (u_m(l),u_i(j)) \x
\prod \Sb j\in \Gm_m(1)\\ l \in \Gm_i(2) \endSb \bt (u_m(j),u_i(l)) \x& \\
\x \prod_{i=1}^N\ \prod_{j=1}^{\la_i}\,u^{-1}_i(j)\x
 F_{\la(1)}(u^{(1)}) \ox F_{\la(2)}(u^{(2)}) \x
\prn1 \prod \Sb j\in \Gm_i(1)\\l \in \Gm_i(2) \endSb
T_{ii}(u_i(l)) \ox T_{i+1,i+1}(u_i(j))&\,.
\endaligned
\endmultline
$$
\endproclaim
\nt
Here the sum is taken over all partitions of the set
$\{(i,j): i=1\lc N,\ j=1\lc \la_i\}$
into disjoint subsets $\Gm(1)$ and $\Gm(2)$,
\vvn-.1>
$$
\alignat2
\Gm_i(l) &=\Gm(l)\cap \{(i,j):j=1\lc \la_i\}\,,\qquad
& \la_i(l) &=\#(\Gm_i(l))\,,\\
u^{(l)} &=\{u_i(j): (i,j)\in \Gm(l)\}\,,
&A(x,y) &=\al(x,y)/\bt (x,y)\,.
\endalignat
$$
\Remark{\kern-.5em}
One can check that the last product
$\prod \prod T_{ii}(u_i(l)) \ox T_{i+1,i+1} (u_i(j))$
\vv.1>
does not depend on ordering of factors as an element of $\Vq\ox\Vq$.
\endremark
\Proof of Theorem {\rm (3.5.1)}.
We will use an induction on $N$. The case $N=0$ is trivial, and the
case $N=1$ is known, \cite{IK}. The general case is similar to the $N=1$ case.
Up to the end of this proof, $R(x,y)$ is the $N^2\x N^2$-matrix corresponding
to $\yq$.
\par
We will use the coproduct formula for $K(x)$
$$
\Dl(K(x))=x^{-1}\bigl(T_{11}(x) \ox K(x)+K(x)\ox M(x)\bigr)
\tag3.5.2
$$
where the matrix product in the second term is supposed,
and will use the commutation relations
$$
\align
T_{11}(x)K(y) &=
A(y,x) K(y) T_{11}(x)-\frac{\gm (y,x)}{\bt(y,x)}K(x)T_{11}(y)\,,
\tagg"3.5.3"\\
\kern13pt M^{(1)}(x) K^{(2)}(y) &=K^{(2)}(y) M^{(1)}(x) R(x,y)/\bt(x,y)
-\frac{\overline{\gm}(x,y)}{\bt(x,y)}K^{(2)}(x)M^{(1)}(y) P \kern-13pt
\tagg"3.5.4"
\endalign
$$
where $P$ is the \perm/ of factors in $\C^N\ox\C^N$.
\par
Take $F_\la(u)$ in the form of (3.4.2). Compute $\Dl(F_\la(u))$.
Substitute in the obtained expression formula (3.5.2) for
$\Dl\bigl(K^{(j)}\pu1j\bigr)$. Pushing factors $T_{11}\pu1j$ and
$M^{(j)}\pu1j$ to the right, using (3.5.3) and (3.5.4), we will have
$$
\align
\kern20pt\Dl (F_{\la(u)}) \approx \sum K^{(\mu_{l+1})}(v(l+1)) \ldots
K^{(\mu_{\la_1})}(v(\la_1)) \ox
K^{(\mu_1)}(v(1)) \ldots K^{(\mu_l)}(v(l))&\kern-20pt
\tagg"3.5.5"\\
\Bigl(T_{11}(w(1))\ldots T_{11}(w(l))\ox
M^{(\mu_{l+1})}(w(l+1))\ldots M^{(\mu_{\la_1})}(w(\la_1))&\x \kern-20pt\\
\x \Phi^{{\mu_1\lc \mu_{\la_1}}}_{v(1)\lc v(\la_1),w(1)\lc w(\la_1)}(u)
\bigl(\Dl (\widetilde{F}_\la(u))\,e\ox\ldots\ox e\bigr)&\Bigr)\,.\kern-20pt
\endalign
$$
Here $\Phi$ is some matrix constructed from the \fn/s $\al(x,y)$, $\bt (x,y)$,
$\gm (x,y)$, and $\gb(x,y)$. Arguments $(w(1)\lc w(l),\ v(\la+1)\lc v(\la_1))$
and $(v(1)\lc v(l),w(l+1)\lc w(\la_1))$ are \perm/s of
$(u_1(1)\lc u_1(\la_1))$.
\par
Say that two terms of this sum corresponding to
$v(1)\lc v(\la_1),w(1)\lc w(\la_1),\mu$
and $v'(1)\lc v'(\la_1),w'(1)\lc w'(\la_1),\mu'$ are {\it similar\/} if
$\{v(1)\lc v(l)\}=\{v'(1)\lc v'(l)\}$ and
$\{v(l+1)\lc v(\la_1)\}=\{v'(l+1)\lc v'(\la_1)\}$ as unordered sequences.
Denote the sum of all similar terms corresponding to given
$v(I)=(v(1)\lc v(l))$ and $v(II)=(v(l+1)\lc v(\la_1))$ by $G(v(I),v(II))$
by $G(v(I),v(II))$. Therefore, (3.5.5) takes the form
$$
\Dl (F_\la(u))\approx \sum_{v(I),v(II)} G(v(I),v(II))\,.
\tag3.5.6
$$
\par
$F_{\la(u)}$ is a symmetric \fn/ of $u_1(1)\lc u_1(\la_1)$. Hence for any
$\theta\in S_{\la_1}$,
$$
 F_\la(u) \approx \bigl (K^{(1)}(u_1(1)) \ldots K^{(\la_1)}(u_1(\la_1))
\bigl (\widetilde{F}_\la(u)\,e\lox e\bigr)\bigr)_\theta\x
\prod_{i=2}^N\ \prod_{j=1}^{\la_i}\,u^{\la_1}_i(j)
\tag3.5.7
$$
where $(\ )_\theta$ means the expression obtained from \rhs/ of (3.4.2)
by a \perm/ $\si$ of \var/s $u_1(1)\lc u_1(\la_1)$.
\par
Now compute the coproduct $\Dl$ of \rhs/ (3.5.7), substitute formula (3.5.2)
in the obtained expression, and rewrite the result in the form (3.5.5).
The sum of all similar terms corresponding to given $v(I)$ and $v(II)$
denote by $G^{\theta}(v(I),v(II))$. As a result, we have
$$
\Dl (F_\la(u)) \approx \sum_{v(I),v(II)} G^{\theta}(v(I),v(II))\,.
\tag3.5.8
$$
\proclaim{(3.5.9) Lemma}
 For any $\theta\in S_{\la_1}$, we have
$$
G^{\theta}(v(I),v(II))=G (v(I),v(II))
$$
for all $v(I)$ and $v(II)$.
\endproclaim
\Proof.
The \YB/ (3.2.1) implies that
$$
R(x,y) (\rho_x\ox\rho_y) \o \Dl'(X)= (\rho_x\ox\rho_y) \o \Dl(X) R(x,y)
\tag3.5.10
$$
for any $X\in\Yq$. We will also use commutation relations
$$
\gather
[T_{11}(x),T_{11}(y)]=0\,,
\tag3.5.11\\
\nn2>
\aligned
\al(x,y)K^{(1)}(x)K^{(2)}(y) &=K^{(2)}(y)K^{(1)}(x)R(x,y)\,,\\
R(x,y)M^{(1)}(x)M^{(2)}(y) &=M^{(2)}(y)M^{(1)}(x)R(x,y)\,.
\endaligned
\endgather
$$
Let $\la_1=2$. We have to compare sums of similar terms resulting from
two \rep/s:
$$
\alignat2
& F_\la(u)=K^{(1)}(u_1(1)) K^{(2)}(u_1(2))
\bigl((\dl \ox\rho_{u_1(1)}\ox \rho_{u_1(2)})\,\o&&
\tagg"3.5.12"\\
&&\llap{$\dsize\Dl^{(2),\si}(F_{\tilde{\la}}\ptu)\,e\ox e \bigr)\x
\prod_{i=2}^N\ \prod_{j=1}^{\la_i}\,u^2_i(j)$}&\\
\nn-8>
\Text{and}
\nn2>
& F_\la(u)=K^{(1)}(u_1(2)) K^{(2)}(u_1(1))
\bigl((\dl \ox\rho_{u_1(2)}\ox\rho_{u_1(1)})\,\o&&
\tagg"3.5.13"\\
&&\llap{$\dsize\Dl^{(2),\si}(F_{\tilde{\la}}\ptu)\,e\ox e \bigr)
\x\prod_{i=2}^N\ \prod_{j=1}^{\la_i}\,u^2_i(j)$}\,.&\\
\nn-8>
\Text{The latter is \eqt/ to}
\nn2>
\kern15pt &F_\la(u)=K^{(2)}(u_1(2)) K^{(1)}(u_1(1))
\bigl(R(u_1(1),u_1(2))\,(\dl \ox\rho_{u_1(1)}\ox\rho_{u_1(2)})\,\o\kern-15pt&&
\tagg"3.5.14"\\
&&\llap{$\dsize\Dl^{(2),\si}(F_{\tilde{\la}}\ptu)\,e\ox e \bigr)\x
\al^{-1}(u_1(1),u_1(2))\,
\prod_{i=2}^N\ \prod_{j=1}^{\la_i}\,u^2_i(j)$}&
\endalignat
$$
(tensor factors $\C^N$ can be interchanged because
the whole expression is scalar; then use (3.5.10)). Set
\vadjust{\vsk-.7>}
$$
\widehat{F}_\la(u)=\Dl\bigl((\dl\ox\rho_{u_1(1)}\ox\rho_{u_1(2)}) \o
\Dl^{(2),\si}(F_{\tilde{\la}}\ptu)\,e\ox e \bigr)\x
\prod_{j=1}^{\la_1}\,u^{-1}_1(j)\x
\prod_{i=2}^N\ \prod_{j=1}^{\la_i}\,u_i(j)\,.
$$
 From (3.5.12) we obtain
$$
\alds
\gather
\aligned
\Dl (F_\la(u))&=\bigl(T_{11}(u_1(1))\ox K^{(1)}(u_1(1))+
K^{(1)}(u_1(1))\ox M^{(1)}(u_1(1)) \bigr) \x \\
&\quad \x \bigl(T_{11}(u_1(2)) \ox K^{(2)}(u_1(2))
+K^{(2)}(u_1(2)) \ox M^{(2)}(u_1(2))\bigr) \bigl(\widehat{F}_\la(u)\bigr)=
\endaligned\\
{\align
=\Bigl(T_{11}(&u_1(1)) T_{11}(u_1(2)) \ox K^{(1)}(u_1(1)) K^{(2)}(u_1(2))+\\
+ K^{(1)} \kern-3pt &\kern3pt
(u_1(1)) K^{(2)}(u_1(2))\ox M^{(1)}(u_1(1)) M^{(2)}(u_1(2))+\\
+ \kern3pt &\kern-3pt
\bigl(A(u_1(2),u_1(1)) K^{(2)}(u_1(2)) T_{11}(u_1(1))-\\
& -\frac{\gm (u_1(2),u_1(1))}{\bt(u_1(2),u_1(1))}
K^{(2)}(u_1(1)) T_{11}(u_1(2))\bigr) \ox K^{(1)}(u_1(1)) M^{(2)}(u_1(2))+
\endalign}\\
\aligned
+K^{(1)}(u_1(1)) T_{11}(u_1(2)) \ox
\bigl(K^{(2)}(u_1(2)) M^{(1)}(u_1(1)) R(u_1(1),u_1(2))/\bt(u_1(1),u_1(2))&-\\
-\frac{\gb(u_1(1),u_1(2))}{\bt(u_1(1),u_1(2))}K^{(2)}(u_1(1))
M^{(1)}(u_1(2))P \bigr)\Bigr)\bigl(\widehat{F}_\la(u)\bigr)&\,,
\endaligned
\endgather
$$
and (3.5.14) leads to
$$
\alds
\gather
\aligned
\Dl (F_\la(u))&=\bigl(T_{11}(u_1(2))\ox K^{(2)}(u_1(2))+K^{(2)}(u_1(2)) \ox
M^{(2)}(u_1(2))\bigr) \x \\
&\quad \x \bigl(T_{11}(u_1(1)) \ox K^{(1)}(u_1(1)) +
K^{(1)}(u_1(1)) \ox M^{(1)}(u_1(1))\bigr) \x \\
&\quad \x \bigl(R(u_1(1),u_1(2)) \widehat{F}_\la(u)\bigr)/\al(u_1(1),u_1(2))=
\endaligned\\
{\align
=\Bigl(T_{11}(&u_1(2)) T_{11}(u_1(1)) \ox K^{(2)}(u_1(2)) K^{(1)}(u_1(1))+\\
+K^{(2)} \kern-2pt &\kern2pt
(u_1(2)) K^{(1)}(u_1(1))\ox M^{(2)}(u_1(2)) M^{(1)}(u_1(1))+
K^{(2)}(u_1(2)) T_{11}(u_1(1))\ox \\
\ox \kern2pt &\kern-2pt
\bigl(K^{(1)}(u_1(1)) M^{(2)}(u_1(2))
P R(u_1(2),u_1(1)) P/ \bt(u_1(2),u_1(1))-\\
&-\frac{\gb(u_1(2),u_1(1))}{\bt(u_1(2),u_1(1))}
K^{(1)}(u_1(2)) M^{(2)}(u_1(1)) P\bigr)+
\endalign}\\
\aligned
+\bigl(A(u_1(1),u_1(2)) K^{(1)}(u_1(1)) T_{11}(u_1(2))-
\frac{\gm (u_1(1),u_1(2))}{\bt (u_1(1),u_1(2))}
K^{(1)}(u_1(2)) T_{11}(u_1(1)) &\ox\\
\ox K^{(2)}(u_1(2)) M^{(1)}(u_1(1)\bigr)\Bigr)
\bigl(R(u_1(1),u_1(2)) \widehat{F}_\la(u)\bigr)/ \al(u_1(1),u_1(2)) &\,.
\endaligned
\endgather
$$
These two sums coincide term by term except terms containing $\gm (x,y)$ or
$\gb(x,y)$. It is due to (3.5.11) and the identity
$$
PR(x,y)PR(y,x)=\al(x,y)\,\al(y,x)
$$
following from the explicit form of the \Rm/. In terms containing $\gm(x,y)$
or $\gb(x,y)$ we move the \perm/ $P$ from right to left. Then they cancel
each other in each sum separately.
\par
The proof for general $\la_1$ can be done as follows. First we reduce
the problem to the case where $\theta$ is a simple \perm/. Then
the latter case is proved similarly to the example $\la_1=2$ given above.
\endproof
\proclaim{(3.5.15) Lemma}
If $\,G(v(I),v(II)) \neq 0$, then
$\{v(1)\lc v(\la_1) \}=\{u_1(1)\lc u_1(\la_1)\}$.
\endproclaim
\Proof.
The factor $K(u_1(1))$ stands in (3.4.2) on the very left place. Hence,
for each term $G(v(I),v(II))$, we have $u_1(1)\in v(I)\cup v(II)$.
Hence, by Lemma (3.5.9), we have $u_1(j)\in \{v(1)\lc v(\la_1)\}$
for any $j=1\lc \la_1$.
\endproof
Compute $G(v(I),v(II))$ where $v(j)=u_1(j)$, $j=1\lc \la_1$. First,
the sum of similar terms $G(v(I),v(II))$ contains exactly one similar term.
To compute this term one must write explicitly $\Dl(F_\la(u))$, take
the monomial
$$
\align
&T_{11}(u_1(1))\ldots T_{11}(u_1(l))
K^{(l+1)}(u_1(l+1))\ldots K^{(\la_1)}(u_1(\la_1))\ox\\
&\ox K^{(1)}(u_1(1))\ldots K^{(l)}(u_1(l))
M^{(l+1)}(u_1(l+1))\ldots M^{(\la_1)}(u_1(\la_1))\,
\bigl(\widetilde{F}_\la(u) e\lox e\bigr)\,,
\endalign
$$
interchange $T_{11}$'s and $K^{(j)}$'s applying (3.5.3), and each time
applying (3.5.3) keep only the first term of \rhs/ of (3.5.3). Then
$$
\align
G(v(I),v(II&))=\prod^l_{m=1}\ \prod^{\la_1}_{j=l+1} A (u_1(j),u_1(m))\x
\prod_{i=1}^N\prod_{j=1}^{\la_i}\,u^{-1}_i(j)\x
\prod_{i=2}^N\ \prod_{j=1}^{\la_i}\,u^{\la_1}_i(j)\x \\
\x& K^{(l+1)}(u_1(l+1))\ldots K^{(\la_1)}(u_1(\la_1))
\ox K^{(1)}) (u_1(1))\ldots K^{(l)}(u_1(l)) \\
\ald
&\;\aligned
\bigl((T_{11}(u_1(1))\ldots T_{11}(u_1(l))\ox
M^{(l+1)}(u_1(l+1))\ldots M^{(\la_1)}(u_1(\la_1)) \x& \\
\x \Dl\bigl(\widetilde{F}_\la(u)\bigr)\,e\lox e\bigr) &\,.
\endaligned
\endalign
$$
Defining relations (3.2.4) are \eqt/ to
$$
\gather
T^{(2)}(y)\,(\id\,\ox\rho^{(2)}_y)\o\Dl(T^{(1)}(x))=
(\id\,\ox\rho^{(2)}_y)\o\Dl' (T^{(1)}(x))\,T^{(2)}(y)
\tag3.5.16\\
\Text{where $\Dl'=P\o\Dl$ and $P$ is the \perm/ of factors. We also have}
\Dl (\widetilde{F}_\la(u))\approx \bigl(\dl\ox\dl\ox\rho^{(1)}_{u_1(1)}\lox
\rho^{(\la_1)}_{u_1(\la_1)}\bigr) \o
\Dl^{(\la_1+1),\hat{\si}}\bigl(F_{\tilde{\la}}(\tilde{u})\bigr)
\tag3.5.17
\vv-.1>
\endgather
$$
where $\hat{\si}(1,2,3\lc \la_1+2)=(1,2,\la_1+2\lc 3)$.
\par
Employing (3.5.16) for $\yq$ and taking into account (3.2.10) and (3.5.17)
we obtain
$$
\alds
\align
&T_{11}(u_1(1))\ldots T_{11}(u_1(l)) \ox
M^{(l+1)}(u_1(l+1))\ldots M^{(\la_1)}(u_1(\la_1))
\Dl\bigl(\widetilde{F}_\la(u)\bigr) \approx \\
&\aligned
\approx &\bigl(T_{11}(u_1(1))\ldots T_{11}(u_1(l)) \ox\id\bigr)
\bigl(\dl\ox\dl\ox\rho^{(1)}_{u_1(1)}\lox \rho^{(\la_1)}_{u_1(\la_1)}\bigr)\o
\Dl^{(\la_1+1),\check{\si}}\bigl(F_{\tilde{\la}}(\tilde{u})\bigr)\x \\
&\x \bigl(\id\ox M^{(l+1)}(u_1(l+1))\ldots M^{(\la_1)}(u_1(\la_1))\bigr)
\approx
\endaligned\\
&\aligned
\approx&
\bigl(\dl\ox\dl\ox\rho^{(1)}_{u_1(1)}\lox\rho^{(\la_1)}_{u_1(\la_1)}\bigr)\o
\Dl^{(\la_1+1), \check{\si}}\bigl(F_{\tilde{\la}}(\tilde{u})\bigr)\x \\
&\x \bigl(T_{11}(u_1(1))\ldots T_{11}(u_1(l))\ox
M^{(l+1)}(u_1(l+1))\ldots M^{(\la_1)}(u_1(\la_1))\bigr)\,,
\endaligned
\endalign
$$
where $\check{\si}(1\lc \la_1+2)=(1,\la_1+2\lc l+3,2,l+2\lc 3)$.
The last equality follows from (3.2.4) for $\Yq$. Using
\vvn-.2>
$$
\gather
\Dl^{(\la_1+1),\check{\si}}=
(\Dl^{(\la_1-l),\si_1}\ox\Dl^{(l),\si_2})\o\Dl\,,\\
\nn1>
\si_1(1\lc \la_1-l+1)=(1,\la_1+2\lc l+3)\,,\\
\si_2(\la_1-l+2\lc \la_1+2)=(2,l+2\lc 3)\,,\\
\nn2> \Text{and} \nn2>
\aligned
M^{(l+1)}(u_1(l+1))\ldots M^{(\la_1)}(u_1(\la_1))\,&e\lox e \approx\\
\approx T_{22}(u_1(l+1))\ldots T_{22}(u_1(\la_1))\,&e\lox e\,,
\endaligned
\vv-.2>
\endgather
$$
we get
\vvn-.5>
$$
\align
G(v(I),v(II&))=
\prod^{l}_{m=1}\ \prod^{\la_1}_{j=l+1}\ A(u_1(j),u_1(m)) \x
\prod_{i=1}^N\prod_{j=1}^{\la_i}\,u^{-1}_i(j)\x
\prod_{i=2}^N\ \prod_{j=1}^{\la_i}\,u^{\la_1}_i(j)\x \\
\x& K^{(l+1)}(u_1(l+1))\ldots K^{(\la_1)}(u_1(\la_1)) \ox
K^{(1)}(u_1(1)) \ldots K^{(l)}(u_1(l)) \x \\
\ald
&\;\bigl((\dl\ox\dl\ox\rho^{(1)}_{u_1(1)}\lox \rho^{(\la_1)}_{u_1(\la_1)})\o
(\Dl^{(\la_1-l)\si_1}\ox\Dl^{(l),\si_2}) \o
\Dl\bigl(F_{\tilde{\la}}(\tilde{u})\bigr) \x \\
&\;\x e\lox e\bigr) \x T_{11}(u_1(1)) \ldots T_{11}(u_1(l)) \ox
T_{22}(u_1(l+1))\ldots T_{22}(u_1(\la_1)) \,.
\endalign
$$
\par
By the induction assumption we substitute \rhs/ of (3.5.1) instead of
$\Dl(F_{\tilde{\la}}(\tilde{u}))$. Finally, using
$$
\rho_y (T_{22}(x))\,e=\al(x,y)\,e,\qquad
\rho_y (T_{ii}(x))\,e=\bt(x,y)\,e\quad \text{for}\ i\ge 3
$$
and Theorem (3.4.2), we obtain
$$
\alds
\multline
\kern-8pt G(v(I),v(II))= \kern-15pt \sum\Sb \Gm (1)\cup\Gm (2)\\
\Gm_1(1)=\{l+1\lc \la_1\}\\ \Gm_2(1)=\{1\lc l\}\endSb
\prod^N_{i=1}\prod\Sb j\in\Gm_i(1)\\l\in\Gm_i(2)\endSb A(u_i(j),u_i(l))\x
\prod^{N-1}_{i=1}\prod\Sb j\in\Gm_i(1)\\ l\in\Gm_{i+1}(2)\endSb
\< A(u_{i+1}(l),u_i(j))\x \\
\aligned
\x\prod_{i=1}^N\prod_{j=1}^{\la_i}\,u^{-1}_i(j)\x
\prod^{N-1}_{i=1} \prod^N_{m=i+1}
\biggl(\prod\Sb j\in\Gm_i(1)\\l\in\Gm_m(2)\endSb \bt (u_m(l),u_i(j)) \x
\prod\Sb j\in\Gm_m(1)\\l\in\Gm_i(2)\endSb \bt (u_m(j),u_i(l))\biggr) \x& \\
 \x F_{\la(1)}(u^{(1)}) \ox F_{\la(2)}(u^{(2)}) \x
\prod^N_{i=1}\prod \Sb j\in\Gm_i(1)\\ l\in\Gm_i(2)\endSb
T_{ii}(u_i(l)) \ox T_{i+1,i+1}(u_i(j))&\,.
\endaligned
\endmultline
$$
All conventions are the same as for (3.5.1).
\par
Due to Lemma (3.5.9) all other terms in (3.5.6) can be obtained from (3.5.18)
permuting $(u_1(1)\lc u_1(\la_1))$. Thus \rhs/ of (3.5.6) is the sum of
expressions (3.5.18) over all partitions $\Gm_1(1)$ and $\Gm_1(2)$ of
$\{1\lc \la_1\}$. Theorem (3.5.1) is proved.
\endproof

\sect (3.6) Weight Functions of a Tensor Product of $\Uq$ Modules
\par
Let $V(1)\lc V(n)$ be \Uqm/s with \hw/s $\La(1)\lc \La(n)$ and \gv/s
$v_1\lc v_n$, \resp/, $z=(z_1\lc z_n)$. Define
$$
\multline
\xi_{\la,V(1)\lc V(n)}(u,z)=\bigl((\phi_{z_1}\lox\phi_{z_n})\o
\Dl^{(n-1)}(F_\la(u))\bigr)\,v_1\lox v_n
\x \prod^N_{i=1}\ \prod_{j=1}^{\la_i}\,u^{n-1}_i(j) \x \\
\nn1>
\x \prod^{N}_{i=1}\,\Bigl(\,[\la_i]_q!\,\prod^{\la_i}_{j=1}\prod^n_{l=1}
\frac{q^{\La_{i+1}(l)} u_i(j)/ {z_l}-q^{-\La_{i+1}(l)}}{q-\q}
\prod^{N+1}_{m=i+1}\ \prod^{\la_i}_{j=1}\ \prod^{\la_m}_{p=1}
\bt(u_m(p),u_i(j))\Bigr)\1
\vv.1>
\endmultline
\tag3.6.1
$$
(cf\.\ (1.3.11) and (3.3.1)). In particular,
\vv-.1>
$$
\xi_{\la,V}(u,y)=(q-\q)^k\,\prod^{N}_{i=1}\ \prod^{\la_i}_{j=1} u_i(j)
\x \eta_{\la,V}(u,y)\,,\qquad k=\sum^{N}_{i=1}\,\la_i\,,
\tag3.6.2
$$
and according to Theorem (3.3.4), $\eta_{\la,V}(u,y)$ is a symmetric \fn/
of $u_i(1)\lc u_i(\la_i)$ for any $i\in \{1\lc N\}$.
\proclaim{(3.6.3) Theorem}
$$
\alds
\multline
\xi_{\la,V(1)\lc V(n)}(u,z)=\prod^{N}_{i=1}\,\Bigl(\,\frac{1}{[\la_i]_q!}\,
\prod^{\la_i}_{j=2}\ \prod^{j-1}_{m=1}\ A(u_i(m),u_i(j)) \Bigr) \x \\
\x \sum\Sb \la(1)+\ldots+\la(n)=\la\\ \si=(\si(1)\lc \si(N))\endSb
\Bigl(\ \prod^n_{j=2}\ \prod^{j-1}_{m=1}\ \prod^{N}_{i=1}
\ \prod^{l_i(j)}_{l=l_i(j-1)+1} C_{i,m}(u_i(\si_l(i)),z_m)\x \\
\x \prod^{N}_{i=1}\ \prod\Sb 1\le a<b\le \la_i\\ \si_a(i)>\si_b(i)\endSb
B(u_i(\si_a(i)),u_i(\si_b(i)))\x \\
\aligned
\x \prod^n_{j=2}\ \prod^{j-1}_{m=1}\ \prod^{N}_{i=2}
\ \prod^{l_i(j)}_{l=l_i(j-1)+1} \ \prod^{l_{i-1}(m)}_{k=l_{i-1}(m-1)+1}
A(u_i(\si_l(i)),u_{i-1}(\si_k(i-1)))\Bigr)\x& \\
\x\,\xi_{\la(1),V(1)}(u^{(1)},z_1)\lox\xi_{\la(n),V(n)}(u^{(n)},z_n)&\,.
\endaligned
\endmultline
$$
\endproclaim
\nt
The sum is taken over all $\la(1)\lc \la(n)\in \Zn$, such that
$\sum^n_{j=1}\la(j)=\la$, and over all \perm/s
$\si\in S_{\la_1}\lx S_{\la_N}$; and we use the notation
$$
\gather
l_i(m)=\sum^{m}_{j=1} \la_i(j)\,,\qquad l_i(0)=0\,, \quad l_i(n)=\la_i\,, \\
\nn-3> \Text{and} \nn-1>
u^{(m)}=\{u_i(j): j\in \si(i)(l_i(m-1)+1\lc l_i(m))\}\,.
\endgather
$$
\Proof.
The following identity is well known:
\vv-.2>
$$
\NN2>
\gather
\sum_{\si\in S_m}\ \prod^{m}_{j=2}\ \prod^{j-1}_{l=1}
\ A(x(\si_l),x(\si_j))=[m]_q!
\tag3.6.4\\
\Text{This is \eqt/ to}
\prod^m_{j=2}\ \prod^{j-1}_{l=1}\ A(x(l),x(j))
\x \sum_{\si\in S_m}\ \prod\Sb 1\le a<b\le m\\ \si_a>\si_b \endSb
B(x(\si_a),x(\si_b))=[m]_q!
\tag3.6.5
\endgather
$$
\par
Decompose the the iterated coproduct in (3.6.1) as
$\Dl^{(n-1)}=(\Dl^{(p-1)}\ox\Dl^{(n-p-1)})\o \Dl $,
use Theorem (3.5.1) to calculate $\Dl (F_\la(u))$ and employ
$$
(\phi_{z_l}\lox\phi_{z_m})\o \Dl^{(m-l)}(T_{ii}(x))\,v_l \lox v_m=
\prod^m_{k=l}\frac{q^{\La_i(k)}x/ {z_k}-q^{-\La_i(k)}}{(q-\q)}
\,v_l \lox v_m\,.
$$
After simple manipulation, we come to
$$
\alds
\multline
\xi_{\la,V(1)\lc V(n)}(u,z)=\sum_{\Gm'\cup\Gm''}
\prod^{N}_{i=1}\ \prod\Sb j\in \Gm'_i\\ l\in \Gm''_i \endSb
A(u_i(j),u_i(l)) \x \\
\aligned
\x \prod^{N-1}_{i=1}\,\prod\Sb j\in \Gm'_i\\ l \in \Gm''_{i+1}\endSb
\< A(u_{i+1}(l),u_i(j))\x
\prod^N_{i=1}\ \prod_{l \in \Gm''_i}\ \prod^p_{k=1}\,C_{i,k}(u_i(l),z_k)\x& \\
\x \prod^{N}_{i=1}\ \frac{[\la'_i]_q!\,[\la_i'']_q!}{[\la_i]_q!}\x
\xi_{\la',V(1)\lc V(p)}(u',z')\ox \xi_{\la'',V(p+1)\lc V(n)}(u'',z'')&
\endaligned
\endmultline
\tag3.6.6
$$
where primes stand for (1) and (2) in (3.5.1) and
$z'=(z_1\lc z_p)$, $z''=(z_{p+1}\lc z_n)$.
\par
To any $\mu \in \Zn$ and $\si\in S_{\la_1}\lx S_{\la_N}$ assign
the set $\Gm^{\mu,\si}=\{u_i(j):j\in \si(i)(1\lc \mu_i)\}$. Say that
$\si,\tilde{\si}$ are $\mu$-\eqt/ if
$\Gm^{\mu,\si}=\Gm^{\mu,\widetilde{\si}}$. We interpret the sum over all
partitions in (3.6.6) as the sum over all $\la',\la''\in \Zn$, such that
$\la'+\la''=\la$, and over all classes of $\la'$-\eqt/ \perm/s.
Replacing $q$-factorials by sums over \perm/s from
\lhs/ of identity (3.6.5) and extracting a common factor we obtain
$$
\alds
\multline
\xi_{\la,V(1)\lc V(n)}=\prod^{N}_{i=1}
\ \Bigl(\,\frac{1}{[\la_i]_q!}\,\prod^{\la_i}_{j=2}\ \prod^{j-1}_{m=1}
A(u_i(m),u_i(j)) \Bigr) \x \\
\x \sum \Sb \la(1)+\ldots+\la(n)=\la\\ \si=(\si(1)\lc \si(N))\endSb
\Bigl( \prod^p_{m=1}\ \prod^{N}_{i=1} \ \prod^{\la_i}_{l=l_i(p)+1}
C_{i,m}(u_i(\si_l (i)),z_m)\x \\
\aligned
\kern-2pt
\x \prod^{N}_{i=1}\ \prod \Sb 1\le a<b\le \la_i\\ \si_a(i)>\si_b(i) \endSb
\kern-5pt B(u_i(\si_a(i)),u_i(\si_b(i)))\x
\prod^{N}_{i=2}\prod^{\la_i}_{l=l_i(p)+1} \ \prod^{l_i(p)}_{k=1}
A(u_i(\si_l (i)),u_{i-1}(\si_k(i-1)))\Bigr)& \x \\
\x\,\xi_{\la',V(1)\lc V(p)}(u',z')\ox \xi_{\la'',V(p+1)\lc V(n)}(u'',z'')&
\endaligned
\endmultline
\tag3.6.7
$$
where $\la'=\sum^p_{m=1}\la(m)$, $\la''=\sum^n_{m=p+1}\la(m)$, and all other
conventions are the same as in (3.6.3). General formula (3.6.3) can be
proved by induction. Identity (3.6.5) must be used in the proof.
\endproof
\proclaim{(3.6.8) Lemma}
Let $w_{\la,V(1)\lc V(n)}(t,z)$ be the \wf/ (1.4.6). Then
$$
\NN1>
\align
\xi_{\la,V(1)\lc V(n)}(t,z)=\prod^{N}_{i=1}\ \Bigl(\,
\frac{(q-\q)^{\la_i}}{[\la_i]_q!}\,\prod^{\la_i}_{i=1}t_i(j)
\ \prod^{\la_i}_{j=2}\ \prod^{j-1}_{m=1}\ A (t_i(m),t_i(j)) \Bigr)\x& \\
\x\,w_{\la,V(1)\lc V(n)}(t,z)&\,.
\endalign
$$
\endproclaim
The lemma follows from (3.6.2), (3.6.3), (3.6.5) and (1.4.4)--\(1.4.6).
\Proof of Theorem {\rm (3.1.1)}.
By definitions (3.2.14) and (3.2.15) of the \Rm/ and definition (3.6.1)
of $\xi_{\la,V(1)\lc V(n)}(u,z)$ we have
$$
R\0_{V(i),V(i+1)}(z_i/z_{i+1})\,\xi_{\la,V(1)\lc V(n)}(t,z)=
P_i\,\xi_{\la,V(1)\lc V(i+1),V(i)\lc V(n)}(t,\tilde{z}_i)\,.
$$
Notation is the same as in (3.1.1). Lemma (3.6.8) completes the proof.
\endproof

\sect (3.7) Rational $R$-matrix
\par
Let $R(x,y)$ be the \Rm/, defined by (1.3.1) with $\al(x,y)=x-y+1$,
$\bt (x,y)=x-y$, $\gm(x,y)=\gb(x,y)=1$, \cite{Y}. It satisfies
the \YB/ (3.2.1). The {\it Yangian} $\Yg$ \cite{D1} is the \aa/ generated
by elements $T^s_{ij}$, $i,j\in \{1\lc N+1\}$, $s=1,2\ldots $ subject to
relations (3.2.4) where $T(x)$ is the $(N+1)\x(N+1)$-matrix with entries
$$
T_{ij}(x)=x\,\dl_{ij}+\sum^\8_{s=1}T^s_{ij}\,x^{1-s}.
$$
The generators $T_{ij}^s$ for $i,j=2\lc N+1$ obey exactly the same relations
as the generators of $\yg$. Denote by
$$
\NN1>
\align
\dl: \Yg &\to \yg \\
T_{ij}^s &\mapsto T^s_{i+1,j+1} \\
\nn2> \Text{the corresponding embedding. The map} \nn4>
\theta_y: \Yg &\to \Yg [[y]] \\
T(x) &\mapsto T(x-y) \\
\nn1> \Text{is an embedding of algebras. Here} \nn1>
(x-y)\1&=\sum^\8_{k=0} y^kx^{-k-1}\,. \\
\nn-5> \Text{The map} \nn-3>
\rho_y: \Yg &\to \En[y] \\
T(x) &\mapsto R(x,y)
\endalign
$$
is a homomorphism. The Yangian $\Yg$ admits a natural coproduct (3.2.8).
Let $\Ng$ be the left ideal in $\Yg$ generated by all $T^s_{ij}$ for $i>j$,
and $\Vg=\Yg \lower2pt\hbox{$\big/$} \lower4pt\hbox{$\Ng$}$.
The symbol $\approx$ will be used if an equality takes place in $\Vg$.
$\Ng$ is a two-sided coideal in $\Yg$. Hence the coproduct $\Dl$ induces
a coproduct $\Vg \to \Vg \ox \Vg\,$.
The embedding $\dl$, the map $\rho_y$ and the coproduct $\Dl$ have
properties (3.2.10). There exists a homomorphism
$$
\NN4>
\align
\phi: \Yg &\to U(\gl)\\
\Text{defined by formulas (2.3.1). Set} \nn1>
\phi_y=\phi \o \theta_y: \Yg &\to U(\gl) [y]\,.
\endalign
$$
Let $V_1$ and $V_2$ be $\gl$ \hwm/ with \gv/s $v_1$ and $v_2$, \resp/.
There exists a linear map
$$
\NN4>
\gather
R\0_{V_1,V_2}(x): V_1 \ox V_2 \to V_1 \ox V_2 \\
\Text{such that}
R\0_{V_1,V_2}(x-y)\,(\phi_x\ox\phi_y) \o\Dl (x)=
(\phi_x\ox\phi_y) \o \Dl'(x)\,R\0_{V_1,V_2}(x-y)
\tag3.7.1 \\
\Text{for any $X\in \Yg$ and}
R\0_{V_1,V_2}(x)\,v_1\ox v_2=v_1\ox v_2\,,
\tag3.7.2
\endgather
$$
see \cite{C}. Here $\Dl'=P\cdot \Dl$, where $P$ is the \perm/ of factors.
Moreover $R\0_{V_1,V_2}(x)$ satisfies the \YB/ (2.1.4) and
$$
R\1_{V_2,V_1}(-x)= P\0_{V_1,V_2} R\0_{V_1,V_2}(x) P\0_{V_2,V_1}
\tag3.7.3
$$
where $P\0_{V_i,V_j}: V_i\ox V_j\to V_j\ox V_i$ is the \perm/.
\vv.1>
{\it This \Rm/ is used in the \qKZ/} (2.1.6).
\vv.1>
\Remark{(3.7.4)}
Properties (3.7.1) and (3.7.2) uniquely define
the \Rm/ for irreducible modules $V_1$ and $V_2$.
\endremark
Let $V(1)\lc V(n)$ be $\gl$ \hwm/s. For $i=1\lc N$ let $P_i$ be
the \perm/ of $i$-th and $(i+1)$-th factors, and
$\tilde{z}_i=(z_1\lc z_{i+1},z_i\lc z_n)$.
\proclaim{(3.7.5) Theorem}
$$
R\0_{V(i),V(j)}(z_i-z_{i+1})\,w_{\la,V(1)\lc V(n)}(t,z)=
P_i\,w_{\la,V(1)\lc V(i+1),V(i)\lc V(n)}(t,\tilde{z}_i)\,.
$$
{\rm (Cf. Theorem (3.3.1).)}
\endproclaim
\nt
The proof is completely analogous to the proof of
Theorem (3.1.1) given in (3.1)--\(3.6). We only mention
four important formulas, corresponding to formulas
(3.6.1), (3.6.2), (3.6.4) and (3.6.8), \resp/:
$$
\NN2>
\alds
\align
& \xi_{\la,V(1)\lc V(n)}(u,z) = (\phi_{z_1}\lox\phi_{z_n})\o
\Dl^{(n-1)}(F_\la(u))\,v_1\lox v_n \x \\
\noalign{\cl{$\dsize \x\prod^{N}_{i=1}\,\Bigl(\,\la_i!\,\prod^{\la_i}_{j=1}
\ \prod^n_{l=1} \bigl(u_i(j)-z_l+\La_{i+1}(l) \bigr)
\prod^{N+1}_{m=i+1}\ \prod^{\la_i}_{j=1}\ \prod^{\la_m}_{p=1}
\bt(u_m(p),u_i(j))\Bigr)\1 ,$}}
\nn4>
& \xi_{\la,V}(u,y) = \eta_{\la,V}(u,y)\,, \\
\nn3>
& \sum_{\si\in S_m}\ \prod^{m}_{j=2}\ \prod^{j-1}_{l=1}
\ A(x(\si_l),x(\si_j))=m!\,,\\
& \xi_{\la,V(1)\lc V(n)}(t,z) = \prod^{N}_{i=1}\,\Bigl(\,\frac{1}{\la_i!}\,
\prod^{\la_i}_{j=2}\ \prod^{j-1}_{m=1} \,A(t_i(m),t_i(j)) \Bigr)\,
w_{\la,V(1)\lc V(n)}(t,z)\,.
\endalign
$$
\vsk.3>

\head 4. Proof of Theorem (1.5.2) \endhead
\removelastskip\nobreak\vsk.4>\nobreak
\nt {\bf (4.1)}\enspace
In Section 4 we'll prove
\proclaim{(4.1.1) Theorem}
The \fn/ $\Psi(z)$ defined by (1.5.1) satisfies the first \eq/ of the \qKZ/
{\rm (1.1.13)}:
$$
Z_1\Psi=q^{\al_1} L\0_{V(1)}(\mu)\,R\1_{V(n)V(1)}\Bigl(\frac{z_n}{z_1}\Bigr)
\ldots R\1_{V(2)V(1)}\Bigl(\frac{z_2}{z_1}\Bigr) \Psi\,.
$$
\endproclaim
The statement that $\Psi(z)$ satisfies the other \eq/s of the \qKZ/
for $i=2\lc n$ can be easily reduced to Theorem (4.1.1). The reduction is
standard; see, for example, \cite{V2, (2.6) and (3.5)}.

\sect (4.2) Proof of Theorem (4.1.1)
\par
 Functions $\{\Phi\}$ introduced in (1.4.8) have the
following properties
$$
\alds
\NN2>
\align
&\quad\qquad \Phi_{t_i(j),z_m}(p\,t_i(j),z_m)=C_{i,m}(t_i(j),z_m)
\,\Phi_{t_i(j),z_m}(t_i(j),z_m)\,,
\tagg"4.2.1" \\
&\quad\qquad\Phi_{t_i(j),z_m}(t_i(j),pz_m)
\,C_{i,m}(t_i(j),pz_m)=\Phi_{t_i(j),z_m}(t_i(j),z_m)\,, \\
&\qquad \Phi_{t_i(a),t_i(b)}(pt_i(a),t_i(b))
=B(t_i(b),t_i(a))
\,\Phi_{t_i(a),t_i(b)}(t_i(a),t_i(b))\,, \\
&\qquad \Phi_{t_i(a),t_i(b)}(t_i(a),pt_i(b))
\,B(pt_i(b),t_i(a))=\Phi_{t_i(a),t_i(b)}(t_i(a),t_i(b))\,, \\
& \Phi_{t_i(a),t_{i+1}(b)}(t_i(a),pt_{i+1}(b))=A (t_{i+1}(b),t_i(a))
\,\Phi_{t_i(a),t_{i+1}(b)}(t_i(a),t_{i+1}(b))\,, \\
& \Phi_{t_i(a),t_{i+1}(b)}(pt_i(a),t_{i+1}(b))
\,A (t_{i+1}(b),pt_i(a))=\Phi_{t_i(a),t_{i+1}(b)}(t_i(a),t_{i+1}(b))\,.
\endalign
$$
\par
 Fix $\la(1)\lc \la(n)\in \Zn$
\vv.1>
and $\la(1)\lsym+\la(n)=\la$.
 Fix $\si=(\si(1)\lc \si(n))\in S_{\la_1}\lx S_{\la_n}$.
Consider the $V(1)_{\la(1)}\lox V(n)_{\la(n)}$-valued
\vv.1>
\fn/ $w_\si(t,z)=
\alh w_{\la(1),V(1)\lc \la(n),V(n);\si}(t,z)$ defined by (1.4.4).
Our first goal is to describe the \fn/
$$
Z_1\<\int_{[0,\xi\8]_p}\<\< \Phi(t,z)\,
w_{\la(1),V(1)\lc \la(n),V(n);\si}(t,z)\,\Om\,.
$$
 For $i=1\lc N$, define $\bar{\si}(i) \in S_{\la_i}$ by the rule:
$$
\align
& \bar{\si}_j(i) =\si_{j+\la_1(i)}(i)\qquad
\text{for $j=1,2\lc \la_2(i)+\la_3(i)\lsym+\la_n(i)$}\,,
\tagg"4.2.2" \\
& \bar {\si}_{j+\la_2(i)+\ldots+\la_n(i)}(i)=\si_j(i)\,.
\tagg"4.2.3"
\endalign
$$
 For $\bar{\si}=(\bar{\si}(1)\lc \bar{\si}(n))\in S_{\la_1}\lx S_{\la_n}$,
consider the \fn/
$$
\bar w_{\bar {\si}}=
w_{\la(2),V(2)\lc\la(n),V(n),\la(1),V(1);\bar {\si}}(t,z)\,.
$$
It is a $V(2)_{\la(2)} \lox V(n)_{\la(n)}\ox V(1)_{\la(1)}$-valued \fn/. Let
$$
P : V(2) \lox V(n) \ox V(1) \to V(1)\ox V(2) \lox V(n)
$$
be the linear map defined by the rule
$x_2\lox x_n\ox x_1 \mapsto x_1\ox x_2\lox x_n$ for all $x_i\in V(i)$. Then
$P\bar w_{\bar {\si}}$ is a $V(1)_{\la(1)} \lox V(n)_{\la(n)}$-valued \fn/.
\proclaim{(4.2.4) Lemma}
$$
Z_1\<\int_{[0,\xi\8]_p}\<\< \Phi(t,z)\,w_\si (t,z)\,\Om=
q^{\al_1} L\0_{V(1)}(\mu)P\< \int_{[0,\xi\8]_p}\<\< \Phi(t,z)\,
\bar w_{\bar {\si}}(t,z)\,\Om\,.
$$
\endproclaim
\proclaim{(4.2.5) Corollary}
Let $w_{\la(1),V(1)\lc \la(n),V(n)}$ be the \fn/ defined by (1.4.5). Then
$$
\align
Z_1\< \int_{[0,\xi\8]_p}\<\< &\Phi\,w_{\la(1),V(1)\lc \la(n),V(n)}\,\Om= \\
& = q^{\al_1} L\0_{V(1)}(\mu) P\< \int_{[0,\xi\8]_p}\<\< \Phi\,
w_{\la(2),V(2)\lc \la(n),V(n),\la(1),V(1)}\,\Om\,.
\endalign
$$
\endproclaim
\Proof.
Let $U(\Phi w_{\si})$ be the \fn/ obtained from $\Phi w_{\si}$ by the
transformation of \var/s:
$$
\alignat2
t_i(j)&\mapsto p\,t_i(j)\qquad &&\text{for $i=1\lc N$ and $j=1\lc \la_i(1)$}\,,
\tag4.2.6 \\
t_i(j)&\mapsto \;t_i(j)\qquad
&& \text{for $i=1\lc N$ and $j=\la_i(1)+1\lc \la_i$}\,,\\
z_1 &\mapsto pz_1\,,\\
z_m &\mapsto \;z_m \qquad &&\text{for $m=2\lc n$}\,.
\vv-.6>
\endalignat
$$
Then
\vv-.4>
$$
\NN4>
\gather
Z_1\< \int_{[0,\xi\8]_p}\<\< \Phi\,w_\si\,\Om=
\int_{[0,\xi\8]_p}\<\<U(\Phi w_\si)\,\Om
\tag4.2.7 \\
\Text{by the Stokes formula (1.2.7). By (4.2.1), we have} \nn4>
U(\Phi w_\si)=q^{\al_1} L\0_{V(1)}(\mu) P\,\Phi\bar w_{\bar {\si}}\,.
\tag4.2.8
\vv-.1>
\endgather
$$
This proves the lemma.
\endproof
\Proof of Theorem {\rm(4.1.1)}.
By (3.1.1) and (3.2.16) we have
$$
\align
q^{\al_1} &L\0_{V(1)}(\mu)\,P\,w_{\la,V(2)\lc V(n),V(1)}=
\tagg"4.2.9" \\
&= q^{\al_1} L\0_{V(1)}(\mu)\,R\1_{V(n),V(1)}\Bigl(\frac{z_n}{z_1}\Bigr)\ldots
R\1_{V(2),V(1)}\Bigl(\frac{z_2}{z_1}\Bigr) w_{\la,V(1)\lc V(n)}\,.
\endalign
$$
This equality and Corollary (4.2.5) prove Theorem (4.1.1).
\endproof

\head 5. Asymptotic \sol/s to \qKZ/ and \BA/ \endhead

\sect (5.1) Asymptotic \sol/s to holonomic \deq/s
\par
Let $V$ be a finite-dimensional vector space, $V^\ast$ its dual space,
$\bra\ ,\ \ket$ their pairing, and
$K_i(z_1\lc  z_n;p)=\sum_{s=0}^\8 K_{is}(z_1\lc  z_n)p^s$ where all
$K_{is}(z_1\lc z_n)$ and $K_{i0}\1(z_1\lc z_n)$ are smooth $\E(V)$-valued
\fn/s. Set $Z_i=\exp(p\der_i)$, $\der_i=\der_{z_i}$, $i=1\lc n$, as a \fps/.
Assume that $K_i(z;p)$ obey \cc/s
$$
Z_iK_j(z;p)\cdot K_i(z;p)=Z_jK_i(z;p)\cdot K_j(z;p)
\tag5.1.1
$$
which in particular mean that for any $i,j$
$$
[K_{i0}(z),K_{j0}(z)]=0\,.
\tag5.1.2
$$
Let $w(z)$ be a common \egv/ of $K_{i0}(z)$ with \eva/s $E_i(z)$:
$$
K_{i0}(z)w(z)=E_i(z)w(z)\,.
\tag5.1.3
$$
Say that $w(z)$ is a \sev/ if $V$ is spanned by $w(z)$ and
the sum of $\,\im(K_{i0}(z)-E_i(z))$, $i=1\lc n$.
Consider a holonomic system of formal \deq/s
$$
Z_i\Psi(z;p)=K_i(z;p)\Psi(z;p)\,.
\tag5.1.4
$$
Let $w(z)$ be a \sev/ of $K_{i0}(z)$ with \eva/s $E_i(z)$
such that $w(z)$ and $E_i(z)$ are smooth \fn/s.
\proclaim{(5.1.5) Theorem}
There exists a formal \sol/
$$
\Psi(z;p)=\exp(\tau(z)/p+\al(z))\sum_{k=0}^\8 \Psi_k(z)p^k
$$
of system {\rm (5.1.4)} such that $\Psi_0(z)=w(z)$ and $\tau(z)$, $\al(z)$,
$\Psi_k(z)$, $k\ge1$, are smooth \fn/s. Such a \sol/ is unique modulo
scalar factor of the form $\exp(\bt_{-1}/p)\sum_{k=0}^\8 \bt_kp^k$
independent of $z_1\lc z_n$.
\endproclaim
The proof will be given at the end of the section.
\proclaim{(5.1.6) Lemma}
Let $G_i\in$ End $ V$, $i=1\lc m$, obey conditions
\roster
\item"a)" $[G_i,G_j]=0\ $ for any $i,j\,$;
\item"b)" $\Cap_{i=1}^m\ker G_i=0\,$.
\endroster
Then there exist $\al_1\lc \al_m$ such that
$\sum_{i=1}^m\al_iG_i$ is invertible.
\endproclaim
\Proof.
 For $m=1$ the statement is obvious. Let $m=2$ and
$V_0=\Cup_{k=1}^\8 \ker G_1^k$. There exists a subspace $V_1$ such that
$G_1V_1\sub V_1$ and $V=V_0\plus V_1$. Evidently $G_2V_a\sub V_a$,
$a=0,1\,$, too. It is clear that $G_1\vst{V_0}$ is nilpotent and
$G_1\vst{V_1}$ is invertible in $V_1$. It follows from condition \,a) \,that
$G_2\vst{V_0}$ is invertible in $V_0$. Hence $G_1+\al G_2$ is invertible
for small $\al$. For the general case the statement can
be proved by induction on $m$.
\endproof
\proclaim{(5.1.7) Corollary}
A linear system $G_iv=w_i$, $i=1\lc m$, has a \sol/ if and only if the
\cc/s $G_iw_j=G_jw_i$ hold. The \sol/ is unique.
\endproclaim
\Proof.
Let $\al_1\lc \al_m$ be such that $G=\sum_{i=1}^m\al_iG_i$ is invertible.
Then necessarily $v=G\1\Big(\sum_{i=1}^m\al_iw_i\Big)$ and equalities
$G_iv=w_i$ evidently follow from the \cc/s.
\endproof
\proclaim{(5.1.8) Lemma}
Let $G_i\in$ End $V$, $i=1\lc m$, obey conditions
\roster
\item"a)" $[G_i,G_j]=0\,$,
\item"b)" $V$ is spanned by $V_0=\Cap_{i=1}^m\ker G_i$
and $V'=\sum^m_{i=1}\,\im G_i\,$.
\endroster
Then a linear system $G_iv=w_i$, $i=1\lc m$, has a \sol/ if and only if
$w_i\in V'$ and the \cc/s $G_iw_j=G_jw_i$ hold. The \sol/ $v\in V'$ is unique.
\endproclaim
\Proof.
It follows from Lemma (5.1.6) that we can find $G$ which is a linear
combination of $G_i$, $i=1\lc  m$, such that $\ker G=V_0$ and
$V=V_0\plus V''$, $V''=\im G$. Then $\im G_i=G_iV''=G_i GV=G G_iV\sub V''$,
so $V'=V''$. Let $G_0$ be a projector on $V_0$ along $V'$. The set $G_k$,
$k=0\lc  m$, satisfy conditions of Lemma (5.1.6) and
the statement follows from Corollary (5.1.7).
\endproof
\proclaim{(5.1.9) Corollary}
Let $G_i\in$ End $V$, $i=1\lc  m$, be a commutative family. The following
statements are \eqt/:
\roster
\item"a)" There exists a \sev/ $v$ of $\{G_i\}$ with \eva/s $\{E_i \}$;
\item"b)" There exists a \sev/ $f$ of $\{G_i^\ast \}$ with \eva/s
$\{E_i \}$.
\endroster
Moreover $v$ and $f$ can be chosen as polynomials in $G_i$, $E_i$,
$i=1\lc  m$, and $\bra f,v\ket\neq 0$.
\endproclaim
\Proof.
Let $v$ be a \sev/ of $\{G_i\}$ with \eva/s $\{E_i \}$. Then we can find
$G=\sum_{i=1}^m\al_iG_i$ such that $v$ spans $\ker G$ and
$V=\ker G\plus\im G$. This is \eqt/ to $V^\ast=\ker\G\plus\im\G$ and
$\dim\ker\G=1$. Let $f\in\ker\G$ be its basic vector. Obviously,
$\bra f,g\ket\neq 0$. And it is also clear that $v$ and $f$
can be chosen as polynomials in $G$ (or $\G$).
\endproof
Let again $w(z)$ be a \sev/ of $\{K_{i0}(z)\}$ with \eva/s $\{E_i(z)\}$ and
$\chi(z)$ the corresponding \sev/ of $K_{i0}^\ast(z)$ such that
$\bra\chi(z),w(z)\ket=1$.
\proclaim{(5.1.10) Theorem}
There exist unique formal power series $\psi(z;p)=\sum_{s=0}^\8 \psi_s(z)p^s$
and $d_i(z;p)=\sum_{s=0}^\8 d_{is}(z)p^s$ such that
$\psi_0(z)=w(z)$, $d_{i0}(z)=E_i\1(z)$, $\bra\chi(z),\psi(z;p)\ket$ $=1$, and
$$
Z_i\psi(z;p)=d_i(z;p)K_i(z;p)\psi(z;p)\,.
$$
$\psi_s(z)$ and $d_{is}(z)$ are \dfl/ polynomials
in $w(z)$, $\chi(z)$, $E_i(z)$, $K_{ik}(z)$, $k=1\lc s$.
\endproclaim
\Proof.
 For any formal power series $a(p)=\sum_{s=0}^\8 a_sp^s$ denote
$\{a(p)\}_s=a_s$ and $\{a(p)\}^s=\sum_{k=0}^sa_kp^k$. The zero order term in
$p$ of the system in question is trivially satisfied. The higher order terms
in $p$ of the system read as follows.
$$
\gather
(d_{i0}(z)K_{i0}(z)-1)\psi_{s+1}(z)+d_{i,s+1}K_{i0}(z)w(z)=\psit_{is}(z)\,,
\tag5.1.11 \\
\bra\chi(z),\psi_{s+1}(z)\ket=0
\endgather
$$
where $\psit_{is}(z)=\{Z_i\psis-\dis K_i(z;p)\psis\}_{s+1}$. Then
$d_{i,s+1}(z)=d_{i0}(z)\*\bra\chi(z),\psit_{is}(z)\ket$,
and due to Lemma (5.1.8) $\psi_{s+1}(z)$ can be found uniquely if the \cc/s
$$
(d_{i0}(z)K_{i0}(z)-1)\psit_{js}(z)=(d_{j0}(z)K_{j0}(z)-1)\psit_{is}(z)
\tag5.1.12
$$
hold. Thus to prove the Theorem it suffices to prove (5.1.12).
\par
Let us rewrite conditions (5.1.1) as
\vv-.25>
$$
\NN3>
\gather
[K_i\1(z;p)Z_i,K_j\1(z;p)Z_j]=0 \\
\Text{and use relations} \nn1>
\{K_i\1(z;p)Z_i\psis-\dis\psis\}^s=0\,.\\
\Text{We have} \nn1>
\aligned
\{K_i\1(z;p)Z_i\psis\}^{s+1} &= \{\dis\psis\}^{s+1}\,+ \\
\nn-1>
+\,\{(K_i\1(z;p)Z_i &- \dis)\psis\}_{s+1} p^{s+1}\,,
\endaligned
\vv-.8>
\endgather
$$
then
\vv-.4>
$$
\NN1>
\align
&\{K_i\1(z;p)Z_i(K_j\1(z;p)Z_j\psis) \}^{s+1}= \\
&= \{Z_i\djs\cdot\dis\psis\}^{s+1} \,+\\
\ald
&\quad+ \,\big(d_{j0}(z)\{(K_i\1(z;p)Z_i-\dis)\psis\}_{s+1}\,+\\
&\quad+ \,K_{i0}\1(z)\{(K_j\1(z;p)Z_j-\djs)\psis\}_{s+1}\big)p^{s+1}\,,
\vv-.5>
\endalign
$$
and finally
\vv-.1>
$$
\align
0 &=\{[K_i\1(z;p)Z_i,K_j\1(z;p)Z_j]\psis\}^{s+1}=\\
&= \{(Z_i\djs\cdot\dis-Z_j\dis\cdot\djs)\psis\}^{s+1}+\\
&\quad+ \big((K_{i0}\1(z)-d_{i0}(z))\{(K_j\1(z;p)Z_j-\djs)\psis\}_{s+1}\,-\\
&\quad- (K_{j0}\1(z)-d_{j0}(z))\{(K_i\1(z;p)Z_i-\dis)\psis\}_{s+1}\big)
p^{s+1}=\\
&= \{(Z_i\djs\cdot\djs-Z_j\dis\cdot\djs)\psis\}^{s+1}+\\
&\quad+ K_{i0}\1(z)K_{j0}\1(z)\big((1-d_{i0}(z)K_{i0}(z))
\psit_{j,s}(z)-(1-d_{j0}(z)K_{j0}(z))\psit_{is}(z)\big)p^{s+1}\,.
\endalign
$$
\par
Applying the \fn/al $\chi(z)$ to the last expression above we can see that
both terms in it must vanish separately. Hence \cc/s (5.1.12) are proved.
Moreover $\psi_s(z)$ and $d_{is}(z)$ are explicitly constructed
as \dip/s in $w(z)$, $E_i(z)$, $K_{ik}$, $k=1\lc s$.
\endproof
\proclaim{(5.1.13) Corollary}
$\,Z_id_j(z;p)\cdot d_i(z;p)=Z_jd_i(z;p)\cdot d_j(z;p)\,$.
\endproclaim
\proclaim{(5.1.14) Lemma}
There exists a unique formal power series
$\tau(z;p)=\sum_{s=0}^\8 \tau_s(z)p^s$ such that
$$
Z_i\exp(\tau(z;p)/p)=d_i\1(z;p)\exp(\tau(z;p)/p)\,.
$$
\endproclaim
\Proof.
Let $f(x)=(e^x-1)/x$. $f(p\der_i)$ is invertible in the sense of a formal
power series. Set $g_i=(f(p\der_i))\1$ and $\dti_i(z;p)=-g_i\ln d_i(z;p)$.
Equation (5.1.14) is now \eqt/ to
$$
\der_i\tau(z;p)=\dti_i(z;p)
\tag5.1.15
$$
and (5.1.13) is transformed into the \cc/s for (5.1.15).
Hence the statement is proved.
\endproof
\Proof of Theorem {\rm (5.1.5)}.
It is obvious that a \sol/ to system (5.1.4) can be obtained as follows:
$\tau(z)=\tau_0(z)$, $\al(z)=\tau_1(z)$, and
$$
\sum_{s=0}^\8 \Psi_s(z)p^s=
\exp\Bigl(\,\tsum_{s=0}^\8\tau_{s+1}(z)p^s\Bigr)\,\psi(z;p)
$$
where \rhs/ has to be re-expanded as a \fps/ in $p$.
\endproof
\proclaim{(5.1.16) Corollary}
The following \eq/s hold
$$
\gather
\exp(\der_i\tau(z))=E_i(z)\,,\\
\der_iE_j(z)\cdot E_i(z)=\der_jE_i(z)\cdot E_j(z)\,.
\endgather
$$
\endproclaim
\Proof.
They follow from Theorem (5.1.10), Corollary (5.1.13) and Lemma (5.1.14)
if we consider the zero and the first order terms in $p$.
\endproof
The leading term of the formal \sol/s $\Psi(z;p)$ admits the invariant
description. Set
$\hat{\Psi}(z;p)=\exp(-\tau(z)/p)\Psi(z;p)=\sum_{i=0}^\8\hat{\Psi}_kp^k$
and define a connection $\D_i$ as follows:
$$
\ln\,(K_i\1(z;p)Z_i)=-\ln K_{i0}+p\D_i+O(p^2)\,.
\tag5.1.17
$$
\proclaim{(5.1.18) Lemma}
 For any $\,i=1\lc n \quad
\D_i\hat{\Psi}_0(z) \in \sum_{i=0}^n\,\im(d_{i0}(z)K_{i0}(z)-1)\,$.
\endproclaim
\Proof.
System (5.1.4) means that for any \fps/ $h(z;p)$
$$
K_i\1(z;p)\,Z_i(h(z;p)\hat{\Psi}(z;p))=
\exp\big((\tau(z)-Z_i\tau(z))/p\big)\,Z_ih(z;p)\Pch\,.
\tag5.1.19
$$
The right hand side is equal to
$$
\gather
\exp(\tau(z)/p)\,Z_i(\exp(-\tau(z)/p)h(z;p))\Pch=\\
=\exp(p\der_i-\der_i\tau(z))h(z;p)\Pch=\exp(p\der_i-
\ln E_i(z))h(z;p)\Pch\,.
\endgather
$$
Thus, (5.1.19) implies
$$
\ln (K_i\1(z;p)Z_i)(h(z;p)\hat{\Psi}(z;p))=(p\der_i-\ln E_i)h(z;p)\Pch\,.
$$
The statement follows from the first order in $p$ of the last equality for
$h(z;p)=1$.
\endproof

\sect (5.2) Eigenvectors of \qKZ/-operators
\par
Let $V(1)\lc V(n)$ be $\Uq$ \hwm/s with \gv/s $v_1\lc\alb v_n$, \resp/. Let
$R\0_{V(i),V(j)}$ be the \Rms/ (3.2.13)--\(3.2.15) acting in $V(1)\lox V(n)$
according to convention (1.1.11). For any $m\in \{1\lc n\}$ define
the \qKZ/-operator
$$
\align
K_m(z)&=R\0_{V(m),V(m-1)}\Bigl(\frac{z_m}{z_{m-1}}\Bigr)\ldots
R\0_{V(m),V(1)}\Bigl(\frac{z_m}{z_1}\Bigr)\x
\tagg"5.2.1"\\
&\quad\x q^{\al_m} L\0_{V(m)}(\mu)\,
R\1_{V(n),V(m)}\Bigl(\frac{z_n}{z_m}\Bigr)\ldots
R\1_{V(m+1),V(m)}\Bigl(\frac{z_{m+1}}{z_m}\Bigr)\,.
\endalign
$$
It is the operator in \rhs/ of the \qKZ/ (1.1.13) for $p=1$. We have
$$
\NN4>
\gather
[K_i(z),K_j(z)]=0\qquad \text{for all $i,j$}\,.
\tag 5.2.2 \\
\Text{This follows from}
[R\0_{V(i),V(j)}(x),L\0_{V(i)}(\mu)L\0_{V(j)}(\mu)]=0
\tag5.2.3
\vv-.1>
\endgather
$$
and \cite{FR, Theorem 5.4}.
\par
Let $w_{\la,V(1)\lc  V(n)}(t,z)$ be the \wf/ introduced by (1.4.6).
 For any $i\in \{1\lc N\}$, $j\in \{1\lc \la_i\}$ set
$$
\NN3>
\align
H_{ij}(t,z)&=q^{\mu_{i+1}-\mu_i}\prod^{n}_{m=1}C_{i,m}(t_i(j),z_m)
\,\prod^{\la_i}\Sb l=1\\ l \neq j \endSb \,B(t_i(l),t_i(j)) \x
\tagg"5.2.4" \\
\nn-3>
&\qquad\x \prod^{\la_{i+1}}_{l=1}A\1 (t_{i+1}(l),t_i(j))
\x \prod^{\la_{i-1}}_{l=1}A (t_i(j),t_{i-1}(l))\,. \\
\Text{For any $m\in \{1\lc n\}$, set}
E_m(t,z)&=q^{\al_m+\bra \mu ,\La(m)\ket}
\prod^{N}_{i=1}\ \prod^{\la_i}_{j=1}\ C\1_{i,m}(t_i(j),z_m)
\tagg"5.2.5"
\endalign
$$
where $\bra\mu,\La(m)\ket=\sum^{N+1}_{i=1}\mu_i\La_i(m)$.
 For given $z_1\lc z_n$ the system of \eq/s
$$
H_{ij}(t,z)=1
\tag5.2.6
$$
on \var/s $t_i(j),\ i=1\lc N$, $j=1\lc \la_i$, is called the system of \BAE/s
\cite{KR}.
\proclaim{(5.2.7) Theorem}
Let $t=\{t_i(j)\}$ satisfy the \BAE/s. Then
\alh $w_{\la,V(1)\lc V(n)}(t,z)\in(V(1)\lox V(n))_\la$ is an \egv/
of operators $K_1\lc K_n$:
$$
K_m(z) w_{\la,V(1)\lc V(n)}(t,z)=E_m(t,z)w_{\la,V(1) \lc V(n)}(t,z)\,.
$$
\endproclaim
\Proof.
Set $\,M_m(z)=\prod^{m}_{i=1} K_i(z)\,$.
\proclaim{(5.2.8) Lemma}
$$
\NN4>
\align
M_m(z) &=\prod^{m}_{i=1}q^{\al_i} L\0_{V(i)}(\mu)\,\tilde M_{m}(z) \\
\Text{where $\tilde M_{m}(z)=M_m(z,\,\mu=0,\,\al=0)$, and}
M_n(z) &=\prod^{n}_{i=1} q^{\al_i} L\0_{V(i)}(\mu)\,.
\endalign
$$
\endproclaim
The lemma easily follows from
$$
R\1_{V(j),V(i)}(x\1)= R\0_{V(i),V(j)}(x)\,,
\tag5.2.9
$$
(cf\.\ (1.1.11)), and (5.2.1)--\(5.2.3).
\par
 From (3.2.14) it follows that for any $X\in \Yq$
$$
\align
\tilde M_m (z)\,&(\phi_{z_1}\lox \phi_{z_n})\o\Dl^{(n-1)}(X)=
\tagg"5.2.10"\\
= &(\phi_{z_1}\lox \phi_{z_n})\o\Dl^{(n-1),\si_m}(X) \tilde M_m (z)
\endalign
$$
where $\si_m(1\lc n)=(m+1\lc  n,1\lc m)$.
Due to (3.6.8) and (5.2.2) we can consider $M_m(z)$ and
$\xi_{\la,V(1)\lc V(n)}(t,z)$ instead of $K_m(z)$ and
$w_{\la,V(1)\lc V(n)}(t,z)$. Relation (5.2.10) is \eqt/ to
$$
\align
\tilde M_m(z) &(\phi_{z_1}\lox \phi_{z_n}) \o
(\Dl^{(m-1)}\ox \Dl^{(n-m-1)})\o\Dl(X)=
\tagg"5.2.11"\\
=& (\phi_{z_1}\lox \phi_{z_n})\o(\Dl^{(m-1)}\ox \Dl^{(n-m-1)})\o
\Dl'(X)\tilde M_m(z)
\endalign
$$
where $\Dl'=P\o\Dl$ and $P$ is the \perm/ of factors. Using formula
(3.6.6) for $\xi_{\la,V(1)\lc V(n)}(t,z)$ and formulas
(1.1.6), (5.2.8), (5.2.11), we get
$$
\alds
\multline
M_m(z)\,\xi_{\la,V(1)\lc V(n)}(t,z)=
\prod^m_{k=1}q^{\al_k+\bra \mu ,\La(k) \ket} \x
\sum_{\Gm'\cup\Gm''}\ \prod^{N}_{i=1}\,
\prod\Sb j\in \Gm'_i\\ l \in \Gm''_i \endSb A(u_i(l),u_i(j))\rlap{$\x$} \\
\aligned
\x \prod^{N-1}_{i=1} \prod\Sb j\in \Gm'_{i+1}\\ l \in \Gm''_i\endSb
A(u_{i+1}(j),u_i(l)) \x \prod^{N}_{i=1}\ \prod_{j\in \Gm'_i}
\ \prod_{k=1}^m\,C_{i,k}(u_i(j),z_k) \x& \\
\x \prod^{N}_{i=1}\ \Bigl(q^{(\mu_{i+1}-\mu_i)\la'_i}
\,\frac{[\la_i']_q!\,[\la'']_q!}{[\la_i]_q!} \Bigr) \x
\xi_{\la',V(1)\lc V(m)}(t',z') \ox \xi_{\la'',V(m+1)\lc V(n)}(t'',z'') &
\rlap{\,.}
\endaligned
\endmultline
\tag5.2.12
$$
Comparing (3.6.6) and (5.2.12) we see that
$\xi_{\la,V(1)\lc V(n)}(t,z)$ is an \egv/ if $t$ satisfies the \BAE/s. Namely,
$$
M_m (z)\,\xi_{\la,V(1)\lc V(n)}(t,z)=
\prod^m_{k=1}E_k (t,z)\,\xi_{\la,V(1)\lc V(n)}(t,z)
\tag5.2.13
$$
which is \eqt/ to (5.2.11).
\endproof
\Remark{(5.2.14)}
Comparing (5.2.8) and (5.2.13) we get a corollary,
$$
\prod_{m=1}^n\ \prod^{N}_{i=1}\ \prod_{j=1}^{\la_i}\,C_{i,m}(t_i(j),z_m)=
\prod^{N}_{i=1}\ q^{(\mu_i-\mu_{i+1})\la_i} .
$$
This equality can also be deduced directly from the \BAE/s (5.2.6).
\endremark

\sect (5.3) Bethe-Anzatz Equations and Critical Points
\par
Let $F(x;p)$ be a \fn/ of $x$ and $p$ defined in a
punctured neighborhood of the line $p=1$. Define
$$
D_xF(x)=\lim_{p\to 1}\frac{F(px;p)}{F(x;p)}
\tag5.3.1
$$
if the limit exists. Say that $x$ is a $p$-critical point of the \fn/
$F(x;p)$ if $D_xF(x)=1$.
\Example
Let $F(x;p)=\nu(x) \exp(\tau(x)/(p-1))$ where
$\nu(x)$ and $\tau(x)$ are smooth. Then $D_xF(x)=\exp(x\tau'(x))$ and $x$
is a $p$-critical point if $\exp(x\tau'(x))=1$.
\endexample
Consider the \fn/ $\Phi(t,z;p)$ introduced in (1.4.9) to solve the \qKZ/.
\proclaim{(5.3.2) Lemma}
$$
\align
D_{t_i(j)} \Phi (t,z) &=H_{ij}(t,z)\,, \\
D_{z_m} \Phi (t,z) &=E_m(t,z)\,.
\endalign
$$
\endproclaim
\proclaim{(5.3.3) Corollary}
 For given $z_1\lc z_n$ the \BAE/s coincide with
the $p$-critical point \eq/s for $\Phi(t,z;p)$.
\endproclaim
The dilogarithm \fn/ $\Li(u)$ is defined by
$$
\align
\Li(u) &= -\int^u_{0}\,\frac{\ln (1-t)}{t}\,dt\,.
\tagg"5.3.4" \\
\nn2> \Text{Set} \nn2>
\Psi(x,y) &= \Li(xy)+\frac12\,\ln x\,\ln y\,,
\tagg"5.3.5"
\endalign
$$
and
$$
\align
\kern10pt \tau(t,z) =& \ln\;\!q\,\biggl(
\sum^{n}_{m=1} (\al_m+\bra \mu ,\La(m) \ket)\ln\,z_m
+\sum^{N}_{i=1}\ \sum^{\la_i}_{j=1}\ (\mu_{i+1}-\mu_i) \ln\,t_i(j)\biggr)+
\kern-10pt \tagg"5.3.6" \\
&+ \sum^{n}_{m=1}\ \sum^{N}_{i=1}\ \sum^{\la_i}_{j=1}\ \biggl(
\Psi \Bigl(\frac{t_i(j)}{z_m},q^{2\La_{i+1}(m)} \Bigr)
- \Psi \Bigl(\frac{t_i(j)}{z_m},q^{2\La_{i}(m)} \Bigr) \biggr)+ \\
&+ \sum^{N-1}_{i=1}\ \sum^{\la_i}_{a=1}\ \sum^{\la_{i+1}}_{b=1}\ \biggl(
\Psi \Bigl(\frac{t_{i+1}(b)}{t_i(a)},1\Bigr)
- \Psi \Bigl(\frac{t_{i+1}(b)}{t_i(a)},q^2 \Bigr) \biggr)+ \\
&+ \sum^{N}_{i=1}\ \sum_{1\le a<b\le\la_i} \biggl(
\Psi \Bigl(\frac{t_i(a)}{t_i(b)},q^2 \Bigr)
- \Psi \Bigl(\frac{t_i(a)}{t_i(b)},q^{-2}\Bigr) \biggr)\,.
\endalign
$$
\proclaim{(5.3.7) Lemma}
$$
\align
\exp\Bigl(t_i(j)\,\frac{\der\tau(t,z)}{\der t_i(j)}\Bigr)&=H_{ij}(t,z)\,,\\
\exp\Bigl(z_m \frac{\der\tau(t,z)}{\der z_m}\Bigr)&=E_m(t,z)\,.
\endalign
$$
\endproclaim
Let $t=t(z)$ be a local \sol/ to the \BAE/s
holomorphically depending on $z$. Then
$$
t_i{(j)}\,\frac{\der \tau(t(z),z)}{\der t_i (j)} = 2\pi \sqrt{-1}\ I_{i,j}
\tag5.3.8
$$
where $I_{ij}$ are integers independent of $z$. Set
$$
\align
& \hat{\tau}(z) = \tau(t(z),z)-2\pi \sqrt{-1}\ \sum_{ij} I_{i,j}\ln t_i(j)\,,
\tag5.3.9\\
& \widehat{E}_m(z) = E_m(t(z),z).
\endalign
$$
\proclaim{(5.3.10) Lemma}
$$
\widehat{E}_m(z)=\exp\Bigl(z_m \frac{\der\hat{\tau}(z)}{\der z_m}\Bigr)\,.
$$
\endproclaim
\proclaim{(5.3.11) Corollary}
$$
z_j \frac{\der}{\der z_j}\ln\widehat{E}_m(z)=
z_m\frac{\der}{\der z_m}\widehat{E}_j(z)\,.
$$
\endproclaim
\nt (Cf. Corollary (5.1.16).)
\proclaim{(5.3.12) Theorem}
$$
(p-1)\ln \Phi (t,z;p) \to \tau(t,z),
$$
if $p\to 1$ from below and all arguments of \fn/s $\Li$'s in {\rm (5.3.6)}
belong to $(0,1)$.
\endproclaim
The theorem follows from the following lemma.
\proclaim{(5.3.13) Lemma}
Let $(u,p)_\8$ be defined by (1.4.7) and $u,p\in (0,1)$. Then
$$
\lim_{p\to 1}(p-1) \,\ln (u,p)_\8=\Li(u).
$$
\endproclaim
\Proof.
$$
(p-1) \ln (u,p)_\8=-\sum^\8_{m=0}(p^m-p^{m+1})\,\frac{\ln (1-p^mu)}{p^m}\,
\Li(u)\,.
$$
\endproof

\sect (5.4) Asymptotic \sol/s to \qKZ/. $\Uq$ case
\par
Let $V(1)\lc V(n)$ be $\Uq$ \hwm/s with \gv/s $v_1\lc v_n$, \resp/.
Then the dual spaces $\V(1)\lc \V(n)$ are right $\Uq$ lowest \wt/ modules
with \gv/s $\va_1\lc \va_n$, \resp/, such that $\bra\va_i,v_i\ket=1$.
 For any $m=1\lc n$ define the \qKZ/-operator
$$
\align
K_m(z)&=R\0_{V(m),V(m-1)}\Bigl(\frac{pz_m}{z_{m-1}}\Bigr)\ldots
R\0_{V(m),V(1)}\Bigl(\frac{pz_m}{z_1}\Bigr)\x
\tagg"5.4.1"\\
&\quad\x q^{\al_m}L\0_{V(m)}(\mu)\,R\1_{V(n),V(m)}
\Bigl(\frac{z_n}{z_m}\Bigr)\ldots
R\1_{V(m+1),V(m)}\left(\frac{z_{m+1}}{z_m}\right)\,.
\endalign
$$
It is the operator in \rhs/ of the \qKZ/ (1.1.13).
\par
 For any set $\{t_i(j),\,i=1\lc N,\,j=1\lc \la_i\}$ define a dual \wf/
\alh $\oma$:
$$
\align
\oma=\Bigl((\phi_{z_1}\lox \phi_{z_n})\o
\Dl^{(n-1)} \bigl({\T_{J_0I_0}}(t,1)\bigr)\Bigr)^\ast\,
\va_1\lox \va_n\,&\x
\tag5.4.2\\
\x \prod_{i=1}^N\ \prod_{j=1}^{\la_i}\,t^{n-1}_i(j)
\x \prod_{i=1}^N\ \prod_{j=1}^{\la_i}\biggl(\prod_{l=1}^n
\,\Bigl(\,q^{\La_{i+1}(l)}t_i(j)/z_l-q^{-\La_{i+1}(l)}\Bigr)\x
\prod_{m=1}^{j-1} A(t_i(m),t_i(j)) &\rlap{$\x$} \\
\x \prod_{m=1}^{j-1} \al(t_i(j),t_i(m)) \x
\prod_{m=i+1}^{N+1}\ \prod_{p=1}^{\la_m}\bt(t_m(p),t_i(j))&\rlap{$\biggr)\1$}
\endalign
$$
where $I_0,J_0$ are defined in (1.3.10).
\proclaim{(5.4.3) Lemma}
Let $\{t_i(j),\ i=1\lc N,\ j=1\lc \la_i\}$ satisfy the \BAE/s {\rm (5.2.6)}.
Then $\oma$ is an \egv/ of $K^\ast_m(z;1)$ with \eva/ $E_m(t,z)$
{\rm (5.2.5)}.
\endproclaim
\nt
The proof is completely similar to the proof of Theorem (5.2.7).
\par
Let $\tau(t,z)$ be defined by (5.3.6) and $k=\sum_{i=1}^N\la_i$. Set
$$
D(t,z)= \det\left[\,t_i(j)\,t_l(m)\,
{\der^2\tau(t,z)\over\der t_i(j)\,\der t_l(m)}\,\right]_{k\x k}\,.
\tag5.4.4
$$
\proclaim{(5.4.5) Conjecture}
Let $\{t_i(j),\,i=1\lc  N,\,j=1\lc \la_i\}$ satisfy the \BAE/s {\rm (5.2.6)}.
Then
$$
\multline
\bra\oma,\oml\ket=(-1)^k(q-\q)^{-k}D(t,z)\,\x \\
\x \prod_{i=1}^Nq^{\la_i(\mu_i-\mu_{i+1})}\x \prod_{i=1}^N
\ \prod_{j=1}^{\la_i}\ \prod_{m=1}^{j-1}B(t_i(j),t_i(m)) \x
\prod_{i=1}^{N-1}\ \prod_{j=1}^{\la_i}\ \prod_{m=1}^{\la_{i+1}}
A(t_{i+1}(m),t_i(j))\,.
\endmultline
$$
\endproclaim
\nt This conjecture was proved for $N=1$ in \cite{K}.
\par
Let us assume that $\{t_i(j),\ i=1\lc N,\ j=1\lc \la_i\}$ satisfy
the \BAE/s (5.2.6) and $\oml$ is a \sev/ of $K_m(z;1)$ in the weight space
$V_\la=\bigl(V(1)\lox V(n)\bigr)_\la$.
Then we can apply all results of section (5.1) to this case after
the appropriate change of notations. Namely $V\!$, $z_1\lc z_n$ and $p$ there
correspond to the weight space $V_\la$, $\ln z_1\lc \ln z_n$ and $\ln p$ here.

\sect (5.5) \qKZ/-operators for $\gl$ Case
\par
The results of (5.2)--\(5.4) have direct analogues for the $\gl$ case.
\par
Let $V(1)\lc  V(n)$ be $\gl$ \hwm/s with \gv/s $v_1\lc v_n$, \resp/.
Let $R\0_{V(i),V(j)}$ be the \Rms/ (3.7.1), (3.7.2) acting in $V(1)\lox V(n)$
according to convention (1.1.11). For any $m\in \{1\lc n\}$ define
the \qKZ/-operator
$$
\multline
K_m(z)=R\0_{V(m),V(m-1)}(z_m-z_{m-1})\ldots R\0_{V(m),V(1)}(z_m-z_1)\x \\
\x \exp(\al_m)\,L\0_{V(m)}(\mu)\,R\1_{V(n),V(m)}(z_n-z_m) \ldots
R\1_{V(m+1),V(m)}(z_{m+1}-z_m)\,.
\endmultline
\tag5.5.1
$$
It is the operator in \rhs/ of the \qKZ/ (2.1.6) for $p=0$. As for
the $\Uq$ case, we have
$$
[K_i(z),K_j(z)]=0\qquad \text{for all $i,j$}\,.
\tag5.5.2
$$
Let $w_{\la,V(1)\lc  V(n)}(t,z)$ be the \wf/ for the $\gl$ case. For any
$i\in \{1\lc N\}$, $j\in \{1\lc \la_i\}$, $m\in \{1\lc n\}$ define the \fn/s
$H_{ij}(t,z)$ and $E_m(t,z)$ by (5.2.4) and (5.2.5), \resp/, replacing their
factors $q^a$ by $\exp(a)$. Then the statement of Theorem (5.2.7) holds.
The proof is completely similar to the proof for the $\Uq$ case.
\par
Let $F(x;p)$ be a \fn/ of $x$ and $p$ defined in a
punctured neighborhood of the line $p=0$. Define
$$
D_x F(x) =\lim_{p\to 0} \frac{F(x+p;p)}{F(x;p)}
\tag5.5.3
$$
if the limit exists. Say that $x$ is a $p$-critical
point of the \fn/ $F(x;p)$ if $D_x F(x)=1$.
\Example
Let $F(x;p)=\nu(x)\exp(\tau(x)/p)$ where $\nu(x)$ and
$\tau(x)$ are smooth. Then $D_x F(x)=\exp(\tau'(x))$, and $x$ is
a $p$-critical point if $\exp(\tau'(x))=1$.
\endexample
Consider the \fn/ $\Phi (t,z;p)$ introduced in (2.4.3) to solve the \qKZ/.
Then the statements of Lemma (5.3.2) and Corollary (5.3.3) hold. Set
$$
\Psi(x)=x\ln\,x
\tag5.5.4
$$
and
\vv-.8>
$$
\align
\kern17pt
\tau(t,z)=& \sum^{n}_{m=1}(\al_m+\bra \mu ,\La(m)\ket) z_m+
\sum^{N}_{i=1}\ \sum^{\la_i}_{j=1}\ (\mu_{i+1}-\mu_i)t_i(j)+
\tagg"5.5.5" \\
&+ \sum^{n}_{m=1}\ \sum^{N}_{i=1}\ \sum^{\la_i}_{j=1}
\ \bigl(\Psi (t_i(j)-z_m+\La_i(m))-\Psi (t_i(j)-z_m+\La_{i+1}(m)\bigr)+
\kern-17pt
\\
&+ \sum^{N-1}_{i=1}\ \sum^{\la_i}_{a=1}\ \sum^{\la_{i+1}}_{b=1}
\ \bigl(\Psi (t_{i+1}(b)-t_i(a)+1)-\Psi (t_{i+1}(b)-t_i(a))\bigr)+ \\
&+ \sum^{N}_{i=1}\ \sum_{1\le a<b\le \la_i}
\bigl(\Psi (t_i(a)-t_i(b)-1)-\Psi (t_i(a)-t_i(b)+1)\bigr)\,.
\endalign
$$
\proclaim{(5.5.6) Lemma}
$$
\align
\exp\Bigl(\frac{\der \tau(t,z)}{\der t_i(j)}\Bigr)&=H_{ij}(t,z)\,, \\
\exp\Bigl(\frac{\der \tau(t,z)}{\der z_m} \Bigr)&=E_m(t,z)\,.
\endalign
$$
\endproclaim
Let $t=t(z)$ be a local \sol/ to the \BAE/s holomorphically depending on $z$.
Then
$$
\frac{\der \tau(t(z),z)}{\der t_i(j)} = 2\pi \sqrt{-1}\ I_{ij}
\tag5.5.7
$$
where $I_{ij}$ are integers independent of $z$. Set
$$
\align
& \hat{\tau}(z) = \tau(t(z),z)-2\pi\sqrt{-1}
\ \sum_{ij} I_{ij} t_i(j)\,,
\tag5.5.8\\
& \widehat{E}_m(z)=E_m(t(z),z)\,.
\endalign
$$
\proclaim{(5.5.9) Lemma}
$$
\widehat{E}_m(z)= \exp\Bigl(\frac{\der \hat{\tau}(z)}{\der z_m}\Bigr)\,.
$$
\endproclaim
\proclaim{(5.5.10) Corollary}
$$
\frac{\der}{\der z_j}\ln \widehat{E}_m(z)=
\frac{\der}{\der z_m} \ln \widehat{E}_j(z)\,.
$$
\endproclaim
\nt (Cf. Corollary (5.1.16).)
\proclaim{(5.5.11) Theorem}
If $p\to+0$ and all arguments of \fn/s $\Psi$'s in {\rm (5.5.5)} are
positive, then there exist constants $\phi_0,\phi_1$ independent of
$t,z$ such that
$$
p \ln \Phi (t,z;p) = \phi_0 \ln p+\phi_1+\tau(t,z)+ O(p\ln p)
$$
in the asymptotic sense.
\endproclaim
The theorem follows from the Stirling formula.

\sect (5.6) Asymptotic \sol/s to \qKZ/. $\gl$ case
\par
Let $V(1)\lc  V(n)$ be $\gl$ \hwm/s with \gv/s $v_1\lc  v_n$, \resp/.
Then the dual spaces $\V(1)\lc \V(n)$ are right $\gl$ lowest \wt/ modules
with \gv/s $\va_1\lc \va_n$, \resp/, such that $\bra\va_i,v_i\ket=1$.
 For any $m=1\lc n$ define the \qKZ/-operator
$$
\multline
K_m(z)=R\0_{V(m),V(m-1)}(z_m-z_{m-1}+p)\ldots R\0_{V(m),V(1)}(z_m-z_1+p)\x \\
\x \exp(\al_m)\,L\0_{V(m)}(\mu)\,R\1_{V(n),V(m)}(z_n-z_m) \ldots
R\1_{V(m+1),V(m)}(z_{m+1}-z_m)\,.
\endmultline
\tag5.6.1
$$
It is the operator on \rhs/ of the \qKZ/ (2.1.6).
\par
 For any set $\{t_i(j),\,i=1\lc N,\,j=1\lc \la_i\}$ define a dual \wf/
\alh $\oma$:
$$
\align
\oma=\Bigl((\phi_{z_1}\lox\phi_{z_n})\o
\Dl^{(n-1)}\bigl(\T_{J_0I_0}(t,0)\bigr)\Bigr)^\ast\,
\va_1\lox \va_n\,&\x
\tag5.6.2\\
\x \prod_{i=1}^N\ \prod_{j=1}^{\la_i}\,t^{n-1}_i(j)
\x\prod_{i=1}^N\ \prod_{j=1}^{\la_i}
\,\Bigl(\,\prod_{l=1}^n\bigl(t_i(j)-z_l+\La_{i+1}(l)\bigr)\x
\prod_{m=1}^{j-1} A(t_i(m),t_i(j)) &\rlap{$\x$} \\
\x \prod_{m=1}^{j-1} \al(t_i(j),t_i(m)) \x
\prod_{m=i+1}^{N+1}\ \prod_{p=1}^{\la_m}\bt(t_m(p), t_i(j)) &\rlap{$\Bigr)\1$}
\endalign
$$
where $I_0,J_0$ are defined in (1.3.10).
\proclaim{(5.6.3) Lemma}
Let $\{t_i(j),\ i=1\lc N,\ j=1\lc \la_i\}$ satisfy the \BAE/s.
Then $\oma$ is an \egv/ of $K^\ast_m(z;0)$ with
\eva/ $E_m(t,z)$ {\rm (5.2.5)}.
\endproclaim
\nt
The proof is completely similar to the proof of Theorem (5.2.7).
\par
Let $\tau(t,z)$ be defined by (5.5.5) and $k=\sum_{i=1}^N\la_i$. Set
$$
D(t,z)=\det
\left[\,{\der^2\tau(t,z)\over\der t_i(j)\,\der t_l(m)}\,\right]_{k\x k}\,.
\tag5.6.4
$$
\proclaim{(5.6.5) Conjecture}
Let $\{t_i(j),\ i=1\lc N,\ j=1\lc \la_i\}$ satisfy the \BAE/s. Then
$$
\multline
\bra\oma,\oml\ket=(-1)^kD(t,z)\,\x \\
\x \prod_{i=1}^N\exp(\la_i(\mu_i-\mu_{i+1}))\x
\prod_{i=1}^N\ \prod_{j=1}^{\la_i}\ \prod_{m=1}^{j-1}B(t_i(j),t_i(m)) \x
\prod_{i=1}^{N-1}\prod_{j=1}^{\la_i}\prod_{m=1}^{\la_{i+1}}
A(t_{i+1}(m),t_i(j))\,.
\endmultline
$$
\endproclaim
This conjecture was proved for $N=1$ in \cite{K} and for $N=2$ and a special
choice of \glm/s in \cite{R1}.
\par
Let us assume that $\{t_i(j),\ i=1\lc N,\ j=1\lc \la_i\}$ satisfy the \BAE/s
and $\oml$ is a \sev/ of operators $K_m(z;1)$ in the \wt/ space
$V_\la=\big(V(1)\lox V(n)\big)_\la$. Then we can apply all results of
section (5.1) to this case taking $V=V_\la$.

\Refs
\widestnumber\key{[JKMO]}

\def\Key#1 #2 //#3 //#4 y#5 v#6 p{\ref \key #1 \by #2 \paper #3 \jour #4
 \yr #5 \vol #6 \pages}

\def\LMP.{Lett. Math. Phys.}
\def\SMD.{Soviet Math. Dokl.}
\def\CMP.{Commun. Math. Phys.}
\def\LMJ.{Leningrad Math. J.}
\def\JPA.{J. Phys. A}
\def\NPB.{Nucl. Phys. B}
\def\IJMP.{Intern. J. Mod. Phys.}

\def\Ku/{Kulish P.P.}
\def\Resh/{Reshetikhin N.Yu.}
\def\Mi/{Miwa T.}
\def\Jim/{Jimbo M.}
\def\Dri/{Drinfeld V.G.}
\def\Kor/{Korepin V.E.}
\def\Var/{Varchenko A.N.}
\def\Fre/{Frenkel I.B.}
\def\Ma/{Matsuo A.}
\def\Ya/{Yang C.N.}

\Key C
Cherednik I.V. //A new interpretation of Gelfand-Zeflin bases
//Duke Math. J. y1987 v54 p563--577 \issue 2
\endref

\Key D1
\Dri/ //Hopf algebras and the quantum \YB/ //\SMD. y1985 v32
p254--258
\endref

\ref \key D2
\by \Dri/ \paper Quasi-Hopf algebras
\jour \LMJ. \yr 1990 \vol 1 \issue 6
\endref

\Key DF
Ding J. and \Fre/ //Isomorphism of two realizations of quantum affine
algebra $U_q(\widehat{\frak{gl}}_N)$ //\CMP. y1993 v156 p277--300 \issue 2
\endref

\Key FR
\Fre/ and \Resh/ //Quantum affine algebras and
holonomic \deq/s //\CMP. y1992 v146 p1--60
\endref

\Key IJ
Idzumi M., Iohara K., \Jim/, \Mi/, Nakashima T. and Tokihiro T //Quantum affine
symmetry in vertex models //\IJMP. y1993 v8 p1479--1511  \issue 8
\endref

\Key IK
Izergin A.G. and \Kor/ //The quantum inverse scattering method
approach to correlation \fn/s //\CMP. y1984 v94 p67--92
\endref

\Key J
\Jim/ //Quantum \Rm/ for the generalized Toda system //\CMP. y1986 v102
p537--547
\endref

\Key JKMO
\Jim/, Kuniba A., \Mi/ and Okado M. //The $A_n^{(1)}$ face models
//\CMP. y1988 v119 p543--565
\endref

\Key Koh
Kohno T. //Monodromy \rep/s of braid groups and \YB/s //Ann. Inst. Fourier
y1987 v37 p139--160
\endref

\Key K
\Kor/ //Calculations of norms of Bethe wave \fn/ //\CMP. y1982 v86
p391--418
\endref

\Key KR
\Ku/ and \Resh/ //Diagonalization of $GL(N)$
invariant transfer-matrices and quantum $N$ waves (Lee model) //\JPA. y1983
v16 pL591--L596
\endref

\Key KRS
\Ku/, \Resh/ and Sklyanin E.K. //\YB/ and \rep/ theory //\LMP. y1981 v5
p393--403
\endref

\Key KZ
Knizhnik V.G. and Zamolodchikov A.B. //Current algebra and Wess-Zumino models
in two dimensions //\NPB. y1984 v247 p83--103
\endref

\Key M1
\Ma/ //\Ji/s of Jordan-Pochhammer type and quantum \KZv/ \eq/s
//\CMP. y1993 v151 p263--273
\endref

\Key M2
\Ma/ //Quantum algebra structure of certain \Ji/s //preprint y1992 v p1--26
\endref

\Key R1
\Resh/ //Calculation of Bethe vector norms for models with $SU(3)$ symmetry
//Zap. Nauch. Semin. LOMI y1986 v150 p196--213 \lang in Russian
\endref

\Key R2
\Resh/ //\J/ type \inl/s, Bethe vectors, and \sol/s to a \dif/ analog of
the \KZv/ system //\LMP. y1992 v26 p153--165
\endref

\Key RS
\Resh/ and Semenov-Tian-Shansky M.A. //Central extensions of quantum current
groups //\LMP. y1990 v19 p133--142
\endref

\ref \key S
\by Smirnov F.A. \book Form factors in completely integrable models
of quantum field theory  \yr 1992\publ World Scientific \publaddr Singapore
\endref

\Key SV
Schechtman V.V. and \Var/ //Arrangements of hyperplanes and Lie algebra
homology //Invent Math. y1991 v106 p139--194
\endref

\Key V
\Var/ //Hypergeometric \fn/s and the \rep/ theory of Lie algebras and
quantum groups //preprint y1992 v p1--401
\endref

\Key V2
\Var/ //Quantized \KZv/ \eq/s, quantum \YB/, and \deq/s for
$q$-hypergeometric \fn/s //preprint y1993 v p1--35
\endref

\Key Y
\Ya/ //Some exact results for the many-body problem in one dimension with
repulsive delta-\fn/ interaction //Phys. Rev. Lett. y1967 v19 p1312--1314
\issue 23
\endref

\Key YY
\Ya/ and Yang C.P. //One dimensional chain of anisotropic spin-spin
interaction {\rm I. Proof of Bethe's hypothesis for ground state in a finite
system} //Phys. Rev. y1966 v150 p321--327
\endref

\endRefs

\enddocument